\documentclass{jfm}
\usepackage[utf8]{inputenc}
\usepackage{graphicx}
\graphicspath{ {Figures/} }
\usepackage{natbib}
\usepackage{amsmath,amssymb,amsfonts,amssymb}
\usepackage{subfig}
\usepackage{setspace}
\usepackage{float}
\usepackage{microtype}
\usepackage[normalem]{ulem}
\usepackage{ragged2e}
\usepackage{xcolor}
\usepackage{tabularx}

\DeclareCaptionJustification{justified}{\justifying}
\usepackage{hyperref}
\hypersetup{
	hyperindex,
	breaklinks,
	colorlinks=true,
	linkcolor=blue,
	citecolor=magenta,
	bookmarks=true,
	bookmarksopen=true,
	bookmarksopenlevel=2,
	pdfstartpage={1},
	pdfstartview={FitH},
	pdfview={FitH 0},%pdfstartview=FitH,pdfview=FitH,%pdfstartview={XYZ null null 1},
	pdfauthor={M.A. Nikolaidis, P. J. Ioannou and B. F. Farrell},
	pdftitle={}}

\usepackage{ifthen}

\def\U{\bm{\mathsf{U}}}

\def\C{\bm{\mathsf{C}}}
\def\I{\bm{\mathsf{I}}}

\def\P{{\bf P}}

\def\A{\bm{\mathsf{A}}}

\def\U{\bm{\mathsf{U}}}

\def\A{{\bf A}}
\def\C{{\bf C}}
\def\I{{\bf I}}

\def\P{{\bf P}}

\definecolor{fgreen}{rgb}{0.0, 0.5, 0.0}
\definecolor{dblue}{rgb}{0.2, 0.2, 0.6}
\definecolor{springgreen}{rgb}{0.09, 0.45, 0.27}
\definecolor{dartmouthgreen}{rgb}{0.05, 0.5, 0.06}
\definecolor{egyptianblue}{rgb}{0.06, 0.2, 0.65}
\definecolor{fireenginered}{rgb}{0.81, 0.09, 0.13}
\definecolor{forestgreen}{rgb}{0.0, 0.27, 0.13}
\definecolor{harvardcrimson}{rgb}{0.79, 0.0, 0.09}
\definecolor{amaranth}{rgb}{0.9, 0.17, 0.31}

\newcommand\redsout{\bgroup\markoverwith{\textcolor{red}{\rule[0.5ex]{2pt}{0.4pt}}}\ULon}

\newcommand{\be}{\begin{equation}}
\newcommand{\ee}{\end{equation}}
\newcommand{\bdm}{\begin{equation*}}
\newcommand{\edm}{\end{equation*}}
\newcommand{\bea}{\begin{eqnarray}}
\newcommand{\eea}{\end{eqnarray}}

\newcommand{\partialf}[2]
{
 \ifthenelse{\equal{#1}{}}{\frac{\partial}{\partial #2}}{\frac{\partial #1}{\partial #2}}
}

\renewcommand{\[}{\left[}
\renewcommand{\]}{\right]}
\newcommand{\<}{\left\langle}
\renewcommand{\>}{\right\rangle}

\newcommand{\df}{\textrm{d}}

\providecommand\bcdot{\boldsymbol{\cdot}}

\newcounter{saveeqn}%

\def\bt{\tilde{\beta}}

\def\st{\sin{\vartheta}}

\def\xv{\mathbf{x}}

\newcommand{\defn}{\ensuremath{\stackrel{\mathrm{def}}{=}}}
\renewcommand{\equiv}{\defn}

\providecommand\bcdot{\boldsymbol{\cdot}}

\newcommand{\ut}{u_\tau}

\renewcommand{\U}{\mathbf{U}}
\renewcommand{\u}{\mathbf{u}}

\shorttitle{Roll-Streak  Dynamics  in Poiseuille Flow Turbulence}
\shortauthor{M.-A. Nikolaidis et al}

\title{Fluctuation  covariance-based study of roll-streak  dynamics in Poiseuille flow turbulence}

\author{Marios-Andreas~Nikolaidis\aff{1},
 Petros J. Ioannou\aff{1,2}
\corresp{\email{pjioannou@phys.uoa.gr}},
 Brian F. Farrell\aff{2}}
\affiliation{\aff{1}Department of Physics, National and Kapodistrian University of Athens, Athens, Greece
\aff{2}Department of Earth and Planetary Sciences, Harvard University, Cambridge, U.S.A. 
}

\begin{document}

\maketitle

\begin{abstract}

Although the roll-streak (R-S) is fundamentally involved in the dynamics of wall-turbulence,   the physical mechanism responsible for its formation and maintenance remains controversial. In this work we investigate the dynamics maintaining the R-S in turbulent Poiseuille flow at  $R=1650$. Spanwise collocation is used to remove spanwise displacement of the streaks and associated flow components, which isolates the streamwise-mean flow R-S component and the second-order statistics of the streamwise-varying fluctuations that are collocated with the R-S. This streamwise-mean/fluctuation partition of the dynamics facilitates exploiting insights gained from the analytic characterization of turbulence in the  second-order statistical state dynamics (SSD), referred to as S3T, and its
closely associated restricted nonlinear dynamics (RNL)  approximation. Symmetry of the statistics 
about the streak centerline permits separation of the fluctuations into sinuous and varicose components. The Reynolds stress forcing induced by the sinuous and varicose fluctuations acting on the R-S is shown to reinforce low- and high-speed streaks respectively. This targeted reinforcement of streaks by the Reynolds stresses occurs continuously as the fluctuation field is strained by the streamwise-mean streak and not intermittently as would be associated with streak-breakdown events. The Reynolds stresses maintaining the streamwise-mean roll arise primarily from the dominant POD modes of the fluctuations, which can be identified with the time average structure of optimal perturbations growing on the streak. These results are consistent with a universal process of R-S growth and maintenance in turbulent shear flow arising from roll forcing generated by straining turbulent fluctuations, which was identified using the S3T SSD.

\end{abstract}


\maketitle

%% SECTION NAMES:
%% Should be lower-case except for proper nouns and abbreviations
%% Should not end with a period

%=========================================================================

\section{Introduction}

Although  turbulent flows exhibit  fluctuations indicative of a stochastic process, 
closer analysis reveals elements of underlying order. 
Efforts to identify and analyze  the origin of this underlying  order in turbulence led to
 the introduction of a structure measure, the two-point correlation function, which 
 was originally interpreted to provide an influence distance from measurements of  flow velocities \citep{Taylor-1935}.  Progress 
 in measuring apparati subsequently allowed collection of increasingly resolved data sets and \citet{Lumley-1967} proposed a
 method to identify coherent structures arising in turbulent flows
 making use of two-point spatial correlation in the flow. In tandem with identification of coherent structure arising from advances in 
experimental observations were attempts to provide  a theoretical basis for the emergence of these coherent structures,
a summary of which can be found in the reviews by
 \citet{Cantwell-1981}, \citet{Robinson-1991} and \citet{Jimenez-2018}. Advances in flow visualization 
 provided additional evidence of  coherent structure in turbulent shear  flows not only in the buffer layer but 
including organized large- and very-large- scale motions throughout turbulent shear flows e.g. \citep{Hutchins-Marusic-2007,Hellstrom-etal-2011}.

A prominent component of  the coherent structure observed in turbulent shear flow is the roll-streak structure (R-S).
This coherent structure alone accounts for a significant fraction of the turbulent fluctuation kinetic energy and
 considerable effort has been devoted to identifying the mechanisms forming and maintaining the R-S
 \citep{Benney-1960, Jang-etal-1986, Hall-Smith-1991,Hamilton-etal-1995,Waleffe-1997,Schoppa-Hussain-2002, Flores-Jimenez-2010,Hall-Sherwin-2010,Hwang-Cossu-2010b,Hwang-Cossu-2011,Farrell-Ioannou-2012,Rawat-etal-2015,Cossu-Hwang-2017,Kwon-Jimenez-2021}.
 As a result of these efforts, it became apparent
  that the R-S is an important component of not only the energy bearing structures but also of the dynamics  underlying 
  the maintenance of   wall-turbulence.  
One role of the R-S in supporting turbulence is to 
 transfer  streamwise mean momentum from the  spanwise homogeneous equilibrium flow, which 
 is maintained by external mean pressure or boundary-associated forcing,  to form a  spanwise 
 inhomogeneous streak  in the flow \citep{Ellingsen-Palm-1975,Landahl-1980}.  This streak in turn makes available 
 rapidly growing streamwise and spanwise dependent perturbations that support subsequent 
 energy transfers from the streamwise mean flow to the fluctuation field 
 required to both generate and maintain the turbulent state.  An example of the former being in transition to 
 turbulence \citep{Westin-etal-1994,Brandt-2004} and of the latter the SSP mechanism \citep{Hamilton-etal-1995,Waleffe-1997}.

The fact that the R-S  does not arise as a  modal  instability  when the  Navier-Stokes  equations (NSE)
in velocity variables are linearized 
about the streamwise-mean flow led to the belief that the R-S cannot arise as an unstable mode 
in the NSE.  
Nonetheless, in shear flow the R-S is the optimally growing structure in the NSE expressed in velocity state 
 variables.  This has been  studied in both the time  domain \citep{Butler-Farrell-1992, Reddy-Henningson-1993} 
 and frequency domain \citep{McKeon-Sharma-2010,McKeon-2017}, 
 so that the  occurrence of optimals with R-S form arising from transient growth of fluctuations in  the turbulence provides 
 a  plausible explanation for the common observation of this structure in turbulent shear flows. 
 { However, R-S formation through transient growth produces initial algebraic growth followed by decay in time and, if randomly forced, a stochastic distribution in space
 because transient growth lacks an organizational mechanism that would produce temporal persistence and spatial organization of the R-S.} 
 The ubiquity, persistence and large scale organization of  the R-S in turbulent shear flow despite  lack of a { modal}
R-S formation instability in the traditional NSE formulation
 resulted in attempts to uncover  explanations alternative to transient growth of initial or continuously forced perturbations to explain 
 the formation and maintenance of the R-S.
 Among these mechanisms are    various regeneration  or  
 self-sustaining processes \citep{Jimenez-Moin-1991,
 Hamilton-etal-1995,Waleffe-1997,Jimenez-Pinelli-1999,Schoppa-Hussain-2002,Hall-Sherwin-2010,Deguchi-Hall-2016}. 
  Alternatively, the R-S has been attributed to  unstable exact coherent structures (ECS)
  \citep{Waleffe-2001,Halcrow-etal-2009}.  
  While unstable ECS can resemble R-S's,  these structures can not occur  in the turbulence because the 
  ECS is an exact solution, which clearly cannot lie on the chaotic attractor of the turbulence.  In  
  this work we study the physical mechanism underlying the  formation 
  and maintenance of the R-S, which manifestly does exist in wall-turbulence.
 In fact the R-S's in this study, as well as the preponderance of those in 
 Poiseuille flow turbulence,  are hydrodynamically stable \citep{Schoppa-Hussain-2002} and therefore the 
 formation and
maintenance of these stable R-S's cannot be attributed to the dynamics of algorithmically 
 constructed unstable ECS that do not  lie on the chaotic attractor of the turbulence.

  It is now recognized that the R-S
 can arise from a  modal instability when  the NSE is expressed in cumulant variables and 
linearized  about the streamwise-mean flow associated with a  background of turbulent fluctuations. 
This modal instability had been overlooked because it 
 has analytic expression only when the 
 NSE is written using a statistical state dynamics (SSD) 
 formulation, such  as S3T \citep{Farrell-Ioannou-2012,Farrell-Ioannou-2017-bifur}.   The dynamics of the S3T 
 SSD is closely approximated by the restricted nonlinear equations (RNL),  which allows insights 
 from the essentially complete characterization of the analytical structure of 
wall  turbulence dynamics by   S3T to 
  be transferred to  RNL,  and from RNL to its DNS 
 companion  \citep{Thomas-etal-2014,Farrell-etal-2016-VLSM,Farrell-etal-2016-PTRSA}.
 The crucial choice of dynamical significance in the formulation of both the S3T and  its RNL approximation is 
 to use a partition into streamwise-mean and fluctuations from the streamwise-mean. 
 This particular partition is crucial to gaining insight into turbulence dynamics 
 because it isolates the interaction between 
 these two components, which comprises the fundamental dynamics maintaining and regulating  the turbulent state.
  The success of this partition in maintaining a realistic turbulent state  
 when the associated SSD is closed at second order 
  implies that interaction between 
 the streamwise-mean flow and the covariance of fluctuations from the streamwise-mean  
 suffices
 for understanding  the physical mechanism sustaining and regulating turbulence in shear flow. 
 Analysis of the S3T SSD reveals that  the influence of the fluctuations on the streamwise mean  component 
occurs through the fluctuation Reynolds 
 stresses, which can be obtained from the covariance component of the SSD.

{ In agreement with simulations,
R-S formation through the S3T modal instability produces  initial  exponential growth in time leading through nonlinear equilibration to persistent stable equilibrium R-S 
with coherent  harmonic organization in space \citep{Farrell-Ioannou-2017-bifur}. 
Although  in turbulent Poiseuille flow  the R-S is subject to disruption, the organization mechanism inherent in 
the S3T dynamics 
still results in streamwise extended R-S in Poiseuille flow turbulence, 
while accounting for the observed 
persistence and   harmonic organization  of the R-S  in the less disrupted wide channel 
Couette turbulence \citep{Avsarkisov-etal-2014,Pirozzoli-etal-2014,Lee-Moser-2018}.}
 
%  The Reynolds stresses obtained using covariance estimates from simulations
% of RNL and DNS  turbulence
% reveal that the roll circulation in both of  these systems is maintained by  Reynolds stress 
% torques similar to those previously observed to be maintaining the roll circulation in  S3T SSD turbulence. 
% These Reynolds stress diagnostics  also confirm   that the
% fluctuations produce roll forcing  systematically  collocated with the
%streak in a manner  required to maintain
%the streak by the lift-up process, as was also found to be the case in the S3T SSD \citep{Farrell-Ioannou-2012,Farrell-etal-2016-VLSM}.
%

In this work we build on previous work in which the structure of the mean and fluctuation components of the 
R-S were identified using POD-based methods \citep{Nikolaidis-POD-2023}. 
However, our aim in this work is to address not structure but rather dynamics, specifically, we
 analyze data
obtained from DNS and RNL simulations of turbulent Poiseuille flows
at $R=1650$ concentrating on
diagnosing the dynamical processes responsible for sustaining the R-S.
In our study of structure in \cite{Nikolaidis-POD-2023} we  departed from traditional POD analysis by incorporating into the analysis the recognition 
 that while the streamwise-mean R-S   is an emergent
coherent structure supported by the Reynolds-stresses of the streamwise-varying fluctuations, in turbulence this structure is subject 
to stochastic displacements in the homogeneous spanwise direction. In order to isolate
the R-S structure while refining the convergence of the second order statistical quantities supporting  
 it 
we collocate the spanwise position of the R-S as indicated by  the spanwise position of the spanwise varying  streak.  
This method  is similar in intent to the slicing and centering methods employed by
\cite{Rowley-2000,Cvitanovic-2012,Cvitanovic-etal-2013,Kreilos-etal-2014} and the conditional space-time (POD)  method \citep{Schmidt-etal-2019} used 
recently to obtain small scale structure in turbulent boundary layers \citep{Saxton-Fox-etal-2022}
and also to the method applied  recently in dynamical mode decomposition (DMD) in turbulent Couette and Poiseuille flows \citep{Marensi-etal-2023}. 
Using collocation we obtained in \cite{Nikolaidis-POD-2023} the mean structure
of the  low-speed and high-speed R-S and verified 
that these collocated R-S structures are nearly identical in DNS and RNL and that the 
associated fluctuations and Reynolds stresses
are also compellingly similar. The mean streak was found to be 
perturbation stable in the NSE when the NSE is expressed in standard velocity variables and to be 
mirror-symmetric about the centerline in the spanwise direction. 
This mirror-symmetry  allows separation of the fluctuations about the centerline into  linearly statistically independent  odd and even components.
The fluctuations with symmetric streamwise and wall-normal velocity components and antisymmetric spanwise velocity component
are referred to as sinuous fluctuations ($\cal S$),
while  the fluctuations with antisymmetric  streamwise and 
wall-normal velocity components and symmetric spanwise velocity component are 
referred to as varicose
fluctuations ($\cal V$).  
%{{\color{egyptianblue} 
%Because of the  mirror symmetry  in the spanwise direction about the center 
%of the streak,  S and V structures are statistically independent.}}
{ While both $\cal S$ and $\cal V$ fluctuations are represented in the POD modes of both low and high speed streaks, the  dominant  POD 
 modes of the fluctuations 
 associated with the low-speed streak  in both DNS and RNL comprise $\cal S$ oblique waves
collocated with the streak.} Moreover, these  dominant fluctuation POD modes  have   the average structure
of  white in energy perturbations evolved linearly on the R-S, white in energy perturbations being chosen so that the 
perturbations that dominate the response reflect only the intrinsic  dynamics of the  evolution of the perturbations, 
which is determined by the perturbations with optimal growth. 
This result that the dominant POD modes
of the streak excited white in energy have the same structure as the fluctuation POD modes 
in both DNS and RNL has a compelling interpretation: the background turbulence is being strained 
by the streak to produce the structures required to support that streak via the SSP mechanism
and these structures can be identified with the optimal  perturbations on the streak \citep{Nikolaidis-POD-2023}.

Having identified and characterized  the mean low-speed and high-speed streaks and the streak-collocated  fluctuation fields,
we proceed in this report to study the streamwise-mean Reynolds stresses arising from these fluctuations in order to identify
the dynamical mechanism responsible for sustaining the rolls that give rise through lift-up
to the  streaks in both DNS and RNL.   A motivation  for establishing the correspondence in the physical mechanism of the SSP between  DNS 
and  RNL
is that  RNL shares its dynamical structure with S3T so that establishing   correspondence of the 
SSP in DNS and RNL implies that  the SSP structure and mechanism in DNS is dynamically the same as that in S3T, which is completely characterized, and  therefore
establishing this correspondence is tantamount to achieving an analytic characterization of the SSP
underlying wall turbulence in the DNS.

\section{Model problem and numerical methods}

The data is obtained from a DNS of a pressure driven constant mass-flux plane Poiseuille flow in a 
channel which is doubly periodic  in the streamwise, $x$, and
spanwise, $z$, direction. The velocity field is decomposed into
the streamwise-mean  component $\U=(U,V,W)$ and  fluctuations  from the mean, $\u=(u,v,w)$.
In this decomposition the R-S  is part of  the mean component of the flow
with the streak component   defined as $U_s(y,z,t)= U- \[ U \]$, where the square brackets $\[ \cdot \]\equiv (1/L_z) \int_0^{L_z} \cdot ~dz  $ denote the spanwise average,
and the roll component has velocities components $(0,V,W)$.

The incompressible non-dimensional NSE governing
the channel flow  in this decomposition are
\begin{subequations}
\label{eq:DNS}
\begin{align}
\partial_t\U&+ \U \bcdot \nabla \U  - \Pi(t) \hat{\mathbf{x}} + \nabla P - R^{-1} \Delta \U = - \overline {\u \bcdot \nabla \u}~,
\label{eq:NSm}\\
 \partial_t\u&+   \U \bcdot \nabla \u +
\u \bcdot \nabla \U  + \nabla p-  R^{-1} \Delta  \u
= -(\u \bcdot \nabla \u - \overline{\u \bcdot \nabla \u}  ) ~.
 \label{eq:NSp}\\
&\nabla \bcdot \mathbf{U} = 0~,~~~\nabla \bcdot \mathbf{u} = 0~.\label{eq:NSdiv0}
\end{align}\label{eq:NSE0}\end{subequations}   
 The  pressure gradient $\Pi(t)$ is adjusted in time to maintain  constant mass flux. 
 Lengths have been made nondimensional by  $h$, 
the channel's half-width, velocities by the time-mean
velocity at the center of the channel, $U_c $,
and time by $h/ U_c$. 
Averaging in $x$ is denotedd by $\overline{( \cdot )} $ 
and averaging in time by $\langle . \rangle$.  
No-slip and impermeable boundaries are placed at $y = 0$
and $y = 2$, in the wall-normal variable.
The Reynolds number is $R=  U_c  h / \nu$, with  $\nu$ the kinematic viscosity.

\begin{table}
\begin{center}
%\begin{tabular}
%\caption{}
%$\Ret= \ut h / \nu= 100$ with  $\ut= \sqrt{ \nu \left.\df [U]/\df y\right|_{\rm w}}$ 
%($\left.\df [U]/\df y\right|_{\rm w}$ is the shear at the wall) is the friction 
%velocity.%and $[L_x^+$,$L_z^+]$ is the channel size in wall units.}
%%\footnotesize\rm
%\centering\vspace{.8em}
%\begin{ruletabular}
\begin{tabular}{@{}*{6}{c}}
\break
 Abbreviation  & $[L_x,L_z]/h$ & $[\alpha,\beta]$ &$N_x\times N_z\times N_y$& $R_\tau$& $R$ \\
 NSE100    & $[4\pi,\;\pi]$& [0.5,\;2] &$128\times 63\times 97$&$100.59$&1650   \\
 RNL100  & $[4\pi,\;\pi]$& [0.5,\;2] &$ 16\times 63\times 97$&$93.18$&1650   \\
\end{tabular}
\caption{\label{table:geometry}Simulation parameters. $[L_x,L_z]/h$ is the domain size in the streamwise, spanwise direction. $[\alpha,\beta] = [2\pi/L_x, 2\pi/L_z]$ denote the fundamental wavenumbers in the streamwise,spanwise direction. $N_x$, $N_z$ are the number of Fourier components after dealiasing and $N_y$ is the number of Chebyshev components. $R_\tau= \ut h / \nu$ 
is the Reynolds number of the simulation based on the friction velocity  $\ut= \sqrt{ \nu \left.\df [ U ] /\df y\right|_{\rm w}}$,where $\left.\df [ U ] /\df y\right|_{\rm w}$ is the shear at the wall.}
%(Add $Re_{bulk}$)}
%\end{ruletabular}
\end{center}
\end{table}

\begin{figure} 
 \vspace{1cm}
  \begin{center}
  \includegraphics[width=1.0\textwidth]{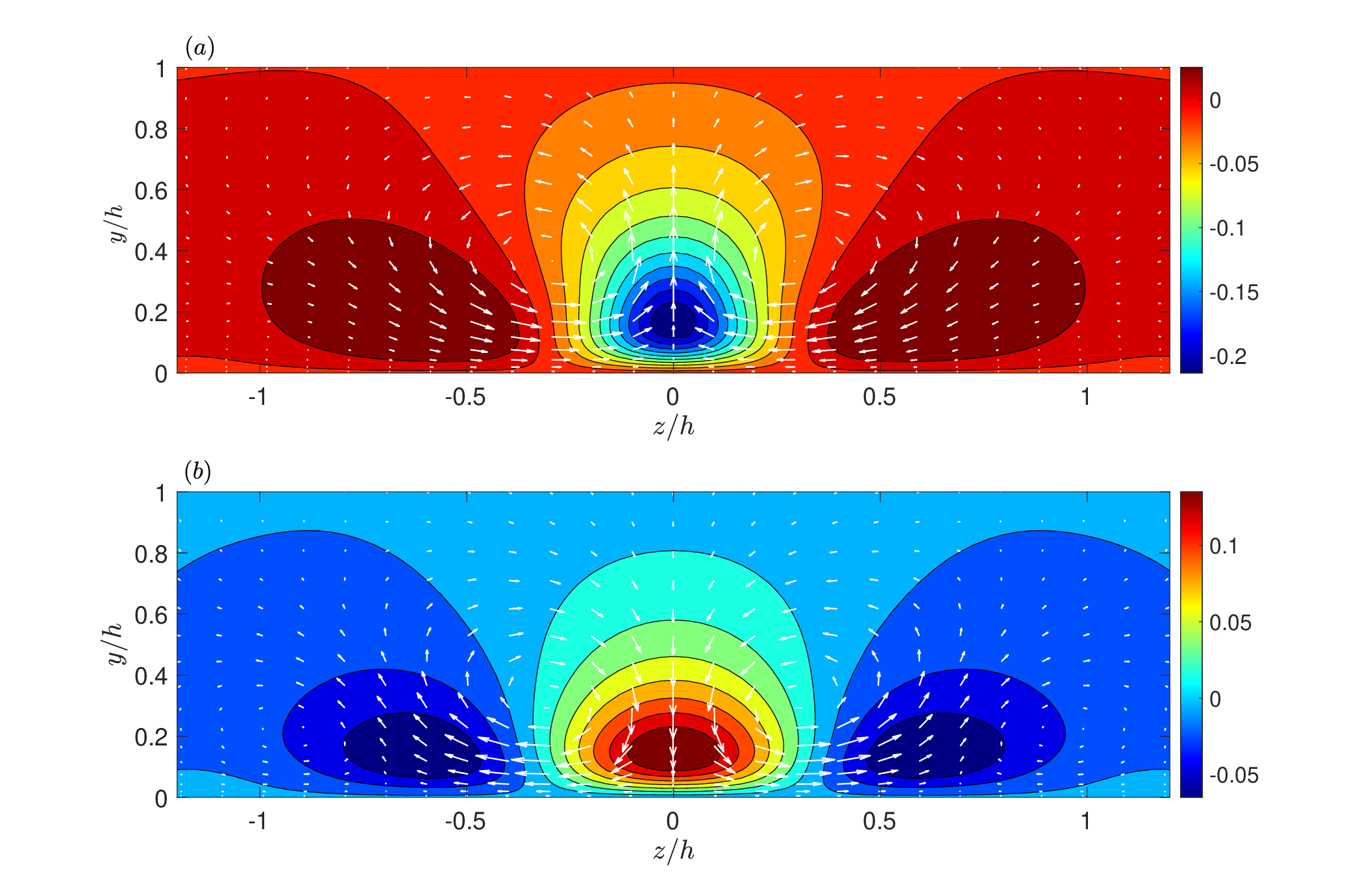}\label{fig:DNS_aligned_avg}  
%    \subfloat[]{\includegraphics[width=0.5\textwidth]{POD3d_inh/U_DNS_avg.eps}\label{fig:DNS_aligned_avg}} % figure 3a
%    \subfloat[]{\includegraphics[width=0.5\textwidth]{POD3d_inh/U_RNL_avg_tuk55.eps}\label{fig:RNL_aligned_avg}}  % figure 3b
 \end{center}
 \caption{Contours of the time-mean collocated  streak, $\langle U_s \rangle$, and vectors  of the roll  
 velocity,  $(\langle W \rangle,\langle V \rangle)$,  for the NSE100  low-speed streak  (a)  and high-speed streak (b).
  The contour interval is  0.025. In (a) the $\max(|\langle U_s \rangle |) = 0.21~U_c$, $\max(\langle V \rangle)=0.024~U_c$. 
  In (b) the  $\max(|\langle U_s \rangle|) = 0.16~U_c$,  $\max(\langle V \rangle)=0.015~U_c$. The contour interval is  $0.025~U_c$.}
  %Panel (b): $\max(|U_s|) = 0.32$, $\max(V)=0.03$.}
 \label{fig:ali2}
\end{figure}

DNS is obtained using NSE  \eqref{eq:DNS}  and for comparison parallel 
simulations are made  with the   RNL approximation of \eqref{eq:DNS}, which 
 is obtained by parameterizing  the fluctuation-fluctuation 
 nonlinearity in equation \eqref{eq:NSp}.
 Except when expressly stated the parameterization used is  
 to set these nonlinear interactions among streamwise non-constant flow components 
in the fluctuation equations  \eqref{eq:NSp} to zero. 
Consequently,  the RNL system of equations is:
 \begin{subequations}
\label{eq:QL}
\begin{align}
\partial_t\U&+ \U \bcdot \nabla \U  - \Pi(t) \hat{\mathbf{x}} + \nabla P - R^{-1}\Delta \U = - \overline {\u \bcdot \nabla \u} ~,
\label{eq:QLm}\\
 \partial_t\u&+   \U \bcdot \nabla \u+
\u \bcdot \nabla \U  + \nabla p-  R^{-1} \Delta  \u
= 0~. 
 \label{eq:QLp}\\
&\nabla \bcdot \mathbf{U} = 0~,~~~\nabla \bcdot \mathbf{u} = 0~.\label{eq:QLdiv0}
\end{align}\label{eq:QLE0}\end{subequations}   
Under this quasi-linear restriction, the fluctuation field interacts nonlinearly  only with the mean, $\U$, flow and  not with itself. This quasi-linear
restriction of the dynamics results in
the spontaneous collapse in the support of the fluctuation field to a 
small subset of streamwise Fourier components, while maintaining conservation of the total flow
energy  $1/2 \int_{\cal D} ~d^3 \xv \left (  |\U|^2 + |\u|^2\right ) $ in the absence of dissipation 
($\cal D$ is  the flow domain).
This restriction in the support of RNL turbulence to a small 
subset of streamwise Fourier components is not imposed but rather is a property of the quasi-linear dynamics.
The fluctuation components retained   by the dynamics  identify the streamwise harmonics that are  energetically  active in
the parametric growth process that sustains the fluctuations \citep{Farrell-Ioannou-2012,Constantinou-etal-2014,Thomas-etal-2014,Thomas-etal-2015,
Farrell-etal-2016-VLSM}.  In a DNS at $R=2250$ these energetically active streamwise harmonics have been shown to synchronize
 the remaining components
 \citep{Nikolaidis-Ioannou-2022}.

The data were obtained from a DNS of Eq. ~\eqref{eq:DNS}, referred to as NSE100,
 and from the associated RNL governed by Eq.~\eqref{eq:QL},
referred to as RNL100. 
The Reynolds number  $R = U_c h/\nu  =1650$ is 
imposed in both the DNS and  the RNL simulations.
A summary of the parameters of the simulations is given in Table~\ref{table:geometry}.
%The mean flow $[ U ]$ and its associated shear   in the NS100 and RNL100 are shown  
%in Fig.~\ref{fig:Mean_flow_Shear} and the time-averaged rms statistics of the fluctuations from the mean flow $[ U ]$, $\u'=\u-[ U ] \hat{x}$ are shown in Fig.\ref{fig:uvw_rms}. Averaging was performed for both simulations on datasets that span approximately $60000~t[ U ]_c/h$ time units.
The RNL100 simulation  is supported by only three streamwise components with wavelengths $\lambda_x/h = 4 \pi, 2 \pi, 4\pi/3 $,
which correspond to the   three lowest  streamwise Fourier components of the channel, $n_x=1,2,3$.

For the  numerical integration the dynamics were expressed in the form of
evolution equations for the wall-normal vorticity and the Laplacian of
the wall-normal velocity, with spatial discretization and Fourier
dealiasing in the two wall-parallel directions and Chebychev polynomials
in the wall-normal direction~\citep{Kim-etal-1987}. Time stepping was
implemented using the third-order semi-implicit Runge-Kutta method.

 \section{Obtaining the streamwise mean R-S and the covariance of the  associated fluctuations using collocation}

In order to analyze the dynamics of the R-S we obtain both the streamwise mean R-S and the time-mean spatial two-point  covariances of the fluctuations
collocated with  the R-S for both  the high-speed and  the low-speed streak. The collocation implementation
is described in \cite{Nikolaidis-POD-2023}.
Briefly the method proceeds by identifying the spanwise location of the streak with the location
of the spanwise coordinate of the $\min(U_s)$ (for low-speed streaks)
and translating the entire flow field  in the spanwise direction  to place the low speed streak minimum at the channel  center $z/h=0$.
{ We have verified  that as the averaging time increases  the time-mean streak approaches  mirror symmetry in the spanwise about the streak centerline.
We enforce this symmetry in the dataset 
and double the available data by symmetrizing  about the aligned streak center.}

The time-mean streak, $\langle U_s \rangle$, obtained from the aligned time-series of  $\min(U_s)$ 
isolates the low-speed streak,
producing a coherent low-speed R-S    at $z/h=0$, while 
away from this core region the
velocity components cancel indicative of their being incoherently correlated with the centered streak.
This collocation procedure is similarly implemented to isolate  the high-speed streak.
The  structures in the $y-z$ plane  of the time-mean low-speed and high-speed R-S in NSE100     are shown in Figs. \ref{fig:ali2}a,b
using contours for $\langle U_s \rangle $ and  vectors for  $(\langle W \rangle ,\langle V \rangle)$. The time-mean flow in the upper region, $y/h>1$,
is to a good approximation spanwise homogeneous (not shown).
The structure of the time-mean streaks in RNL100 are similar (cf. \cite{Nikolaidis-POD-2023}). 
{ It is important to note that although low-speed streaks are associated with flanking high-speed streak components,
the low-speed streaks are isolated structures in the statistical mean because of the  decoherence of the spanwise location of the streaks in Poiseuille flow.
In contrast, simulations of Couette flow turbulence in wide channels reveal that the streaks exhibit long range correlation \citep{Avsarkisov-etal-2014,Pirozzoli-etal-2014,Lee-Moser-2018}.
The implication is that in wide channel Couette flow  a collocation procedure would   not be necessary because the turbulence would exhibit a full array of spanwise periodic  low- and high-speed R-S
rather than the random distribution of isolated R-S seen in Poiseuille flow.}
%a full array of periodic  low- and high-speed R-S is expected.}

\begin{figure} 
 \vspace{1cm}
  \begin{center}
\includegraphics[width=1\textwidth]{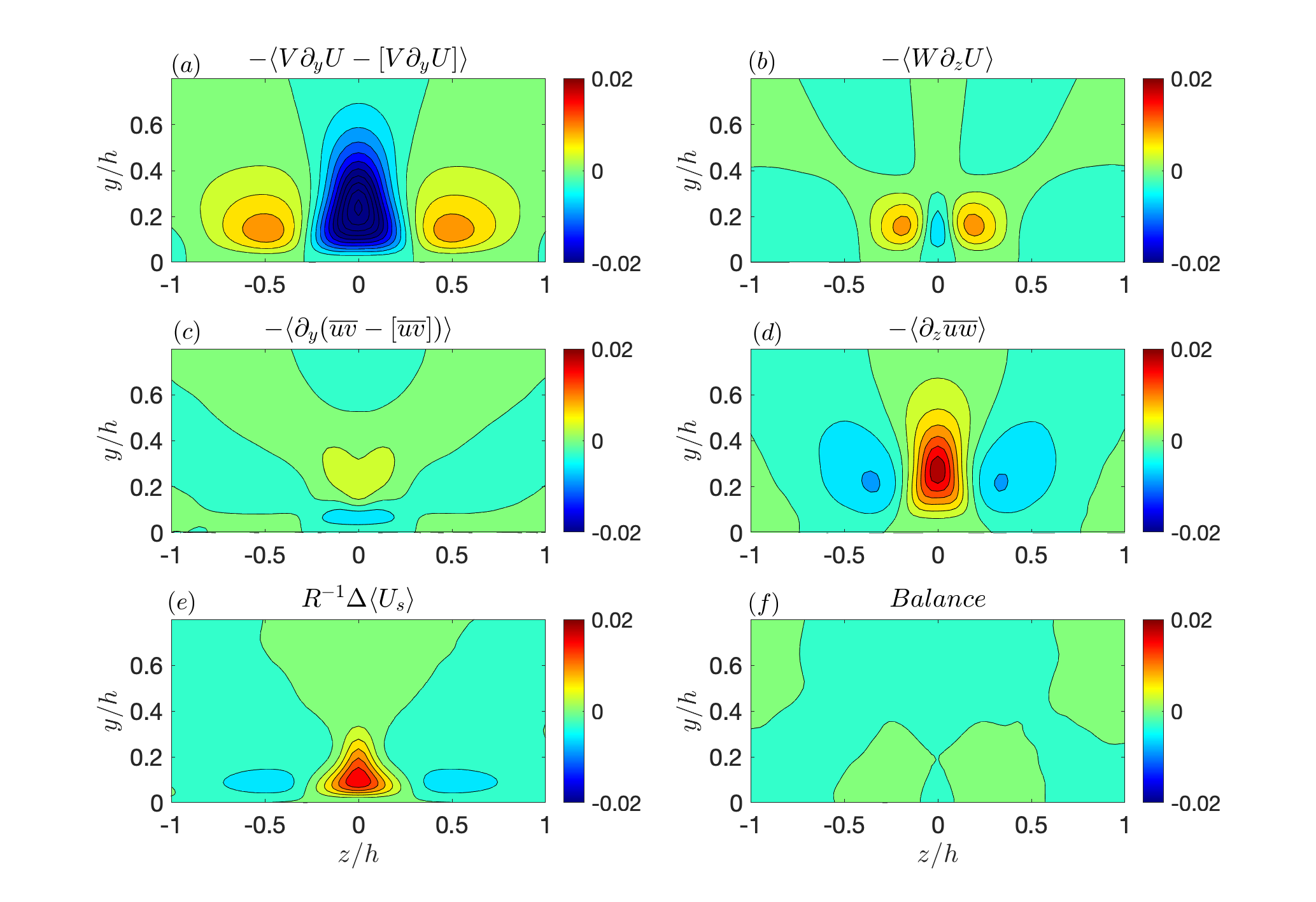} % figure 3a
%    \subfloat[]{\includegraphics[width=0.5\textwidth]{POD3d_inh/balance_stresses_rnl_Ox_full2_notuk.eps}\label{fig:RNL_Ox_balance}}  % figure 3b
 \end{center}
 \caption{For the low-speed streak in NSE100 shown are contours in the $(y,z)$ plane of 
 (a): $ - \langle V \partial_y U - [V \partial_y U] \rangle$, (b): $- \langle W \partial_z U_s \rangle$, 
 (c): $-\langle \partial_y (\overline{u v} -[\overline{uv}])\rangle$, (d): $- \partial_z \langle \overline{u w}  \rangle$,
 and (e):   $R^{-1} \Delta \langle U _s \rangle$.
 The sum shown  in (f)  confirms that the above terms are in balance.  The contour  interval is $0.003~U_c^2/h$.}
 \label{fig:DNS_Us_balance}
\end{figure}

% - \langle V \partial_y (U_s + [U]) \rangle - \langle W \partial_z U_s \rangle -\partial_y \langle \overline{u v} \rangle - \partial_z \langle \overline{u w}  \rangle + R^{-1} \Delta \langle U _s \rangle = 0.
\begin{figure} 
 \vspace{1cm}
  \begin{center}
\includegraphics[width=1\textwidth]{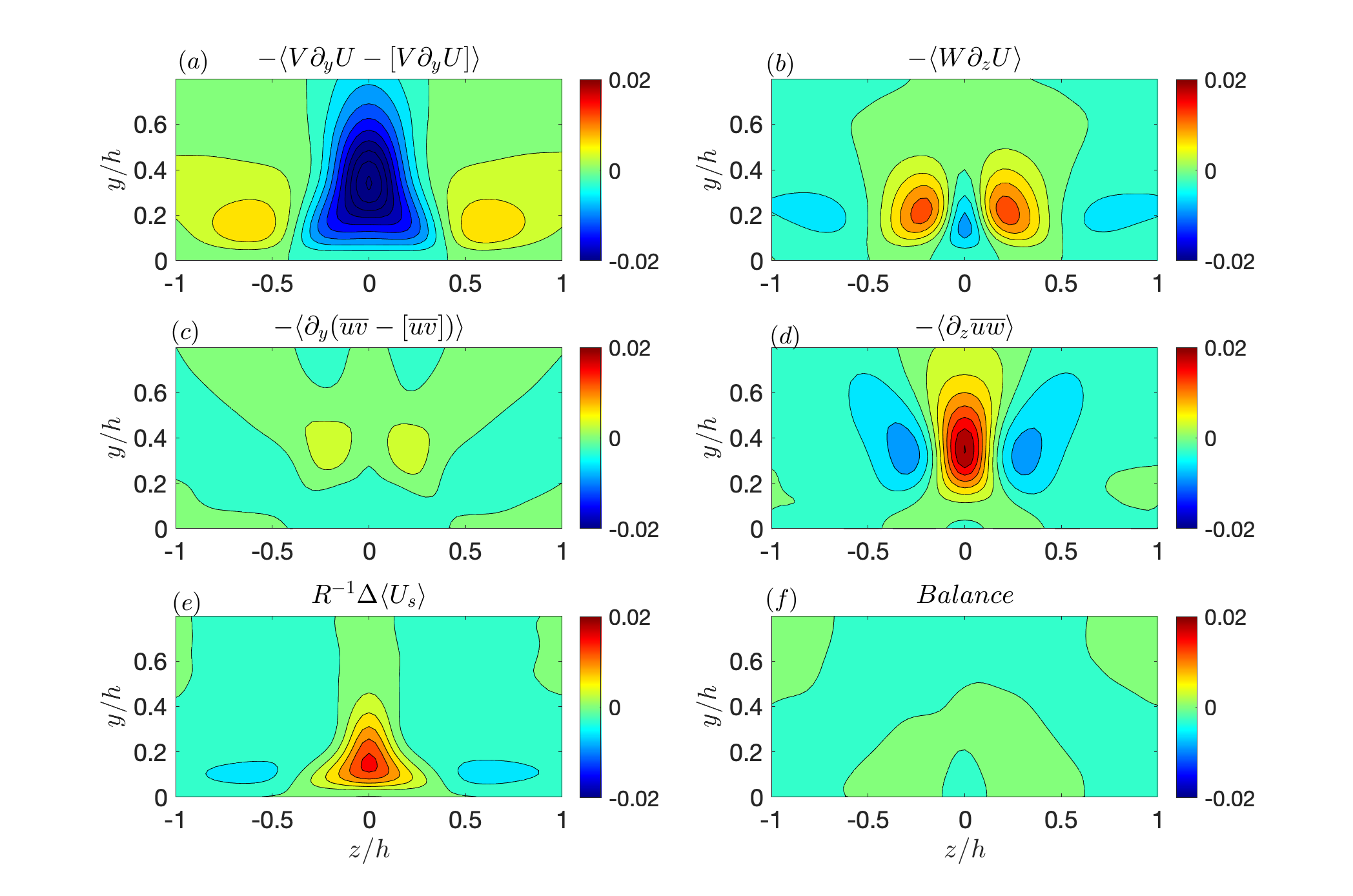} % figure 3a
%    \subfloat[]{\includegraphics[width=0.5\textwidth]{POD3d_inh/balance_stresses_rnl_Ox_full2_notuk.eps}\label{fig:RNL_Ox_balance}}  % figure 3b
 \end{center}
 \caption{As in Fig. \ref{fig:DNS_Us_balance} for RNL100. The contour interval is $0.003~U_c^2/h$. }
 \label{fig:RNL_Us_balance}
\end{figure}

\begin{figure} 
 \vspace{1cm}
  \begin{center}
\includegraphics[width=1\textwidth]{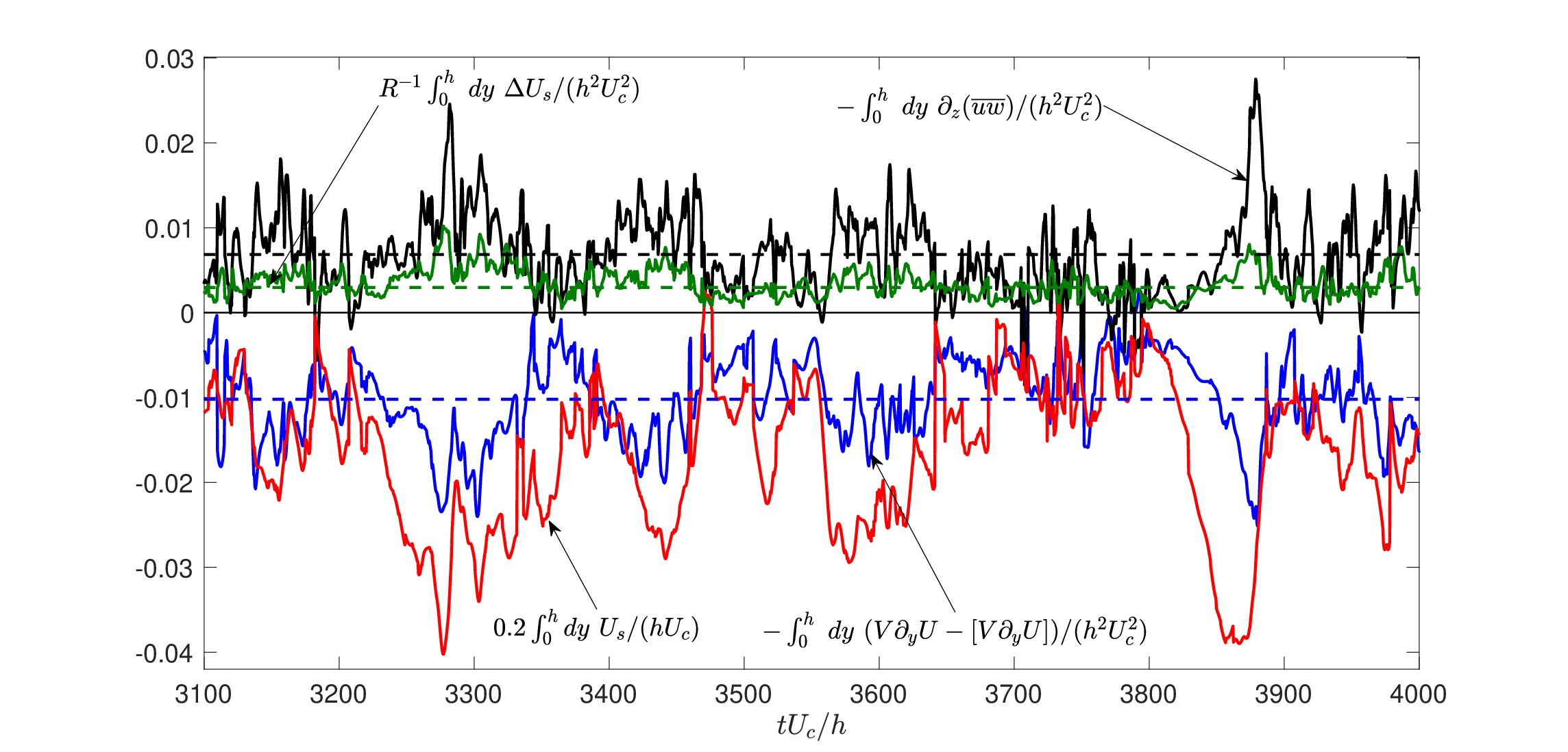} % figure 3a
%    \subfloat[]{\includegraphics[width=0.5\textwidth]{POD3d_inh/balance_stresses_rnl_Ox_full2_notuk.eps}\label{fig:RNL_Ox_balance}}  % figure 3b
 \end{center}
 \caption{Contributions to streak maintenance and regulation in NSE100. The scaled average streak amplitude  at the centerline of the low-speed 
 streak  ($0.2 \int_0^h dy ~U_s /h U_c $) is shown in red. 
 The average  streak acceleration by lift-up  at the centerline of the low-speed streak ($ -\int_0^h dy~ (V \partial_y (U)-[V \partial_y U])/ (h^2 U_c^2)$) (blue)
 is opposed by the acceleration due to diffusion  (green) ($R^{-1} \int_0^h~dy~\Delta U_s/(h^2 U_c^2)$)
 and downgradient momentum transport by  the streamwise varying fluctuations (black) ($ -\int_0^hdy~ \partial_z (\overline{uw})/(h^2 U_c^2)$).
  The dashed lines with the corresponding colors indicate the mean values taken over the entire dataset.  
   This figure shows that maintenance and regulation of the streak is  occurring continuously in time  and  is not
  confined to  bursting events. }
 \label{fig:streak_prod}
\end{figure}

%\begin{figure} 
% \vspace{1cm}
%  \begin{center}
%\includegraphics[width=1\textwidth]{E_comp_dns.eps} % figure 3a
%%    \subfloat[]{\includegraphics[width=0.5\textwidth]{POD3d_inh/balance_stresses_rnl_Ox_full2_notuk.eps}\label{fig:RNL_Ox_balance}}  % figure 3b
% \end{center}
% \caption{
% {\color{egyptianblue} Time-series of streak, roll and perturbation energy in NS100. Shown are the acceleration due to  lift-up  (blue)   
%and the negative of the acceleration  due to Reynolds stress divergence (black) (cf. Fig. \ref{fig:streak_prod}). The time-series have been shifted by the  $1.2 ~h/U_c$ 
%lag between them which was obtained over the entire dataset.
%These two accelerations are highly correlated (correlation coefficient $0.72$) revealing  that a  tight quasi-equilibrium    between lift-up and  
%downgradient momentum transfer characterizes the maintenance and  regulation of the streak amplitude. }}
% \label{fig:streak_prod1}
% \end{figure}

\begin{figure} 
 \vspace{1cm}
  \begin{center}
\includegraphics[width=1\textwidth]{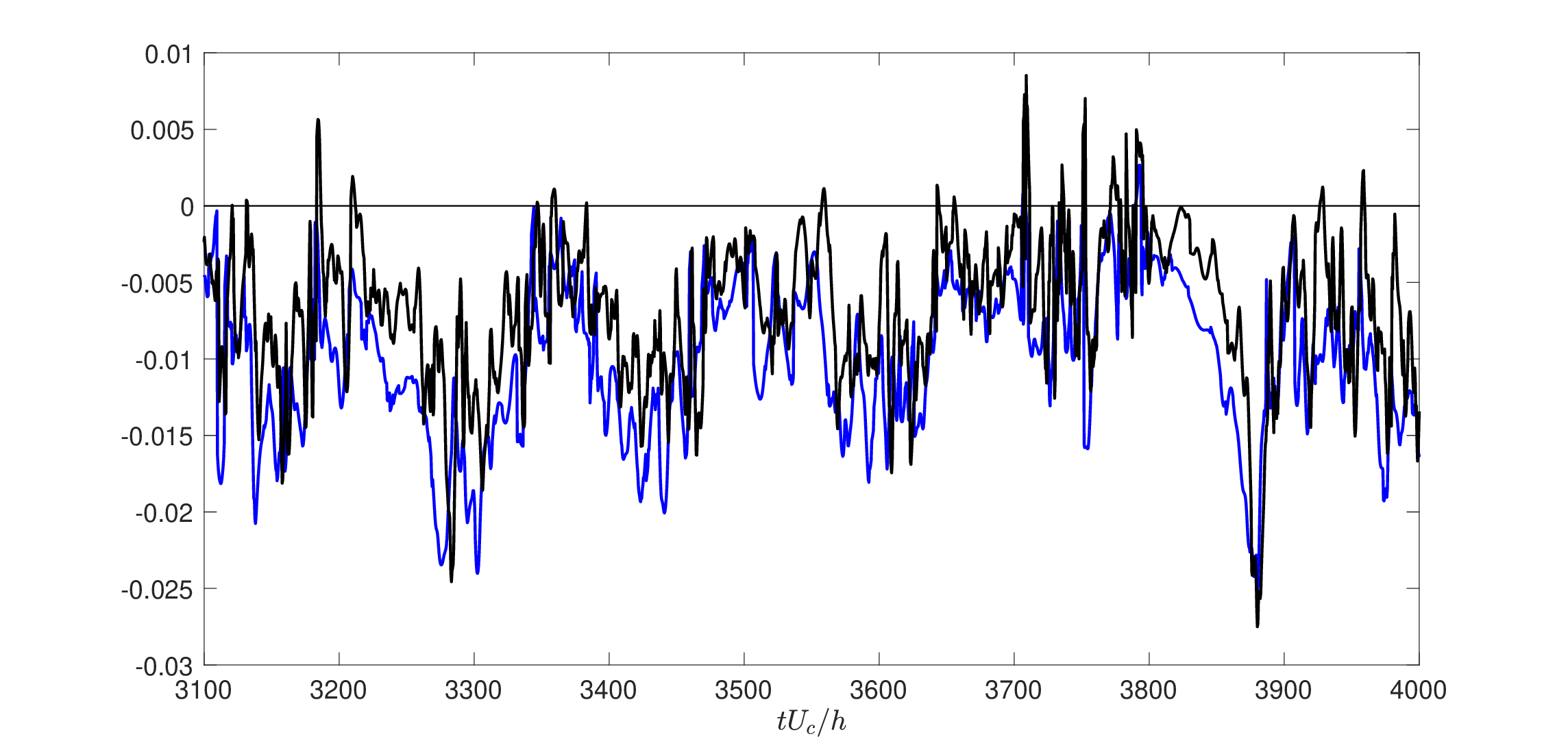} % figure 3a
%    \subfloat[]{\includegraphics[width=0.5\textwidth]{POD3d_inh/balance_stresses_rnl_Ox_full2_notuk.eps}\label{fig:RNL_Ox_balance}}  % figure 3b
 \end{center}
 \caption{Comparison between  the primary components maintaining and regulating the
 low-speed streak in NSE100. Shown are the acceleration due to  lift-up  (blue)   
and the negative of the acceleration  due to Reynolds stress divergence (black) (cf. Fig. \ref{fig:streak_prod}). The time-series have been shifted by the  $1.2 ~h/U_c$ 
lag between them which was obtained over the entire dataset.
These two accelerations are highly correlated (correlation coefficient $0.72$) revealing  that a  tight quasi-equilibrium    between lift-up and  
downgradient momentum transfer characterizes the maintenance and  regulation of the streak amplitude. }
 \label{fig:streak_prod1}
 \end{figure}
%\section{Diagnostic analysis of the covariance of perturbations from the mean roll-streak structure}

Having isolated at each time instant the streamwise mean R-S  with streamwise velocity $ U(y,z,t)$,
wall-normal velocity $V(y,z,t) $ and spanwise velocity $ W(y,z,t) $, we  
Fourier decompose in the streamwise direction the  
fluctuation velocities  collocated with the streak: 
\be
\u= [ u(\xv,t), v(\xv,t), w(\xv,t)    ]^T,
\ee
and calculate the time-mean covariance 
\be
\C_{k_x}(y_1,z_1,y_2,z_2) =\< \u_{k_x}(y_1,z_1) \u_{k_x}^{\dagger}(y_2,z_2) \>~,
\label{eq:Ck}
\ee
where ${\u}_{k_x}(y_i,z_i))$
is the  amplitude of the  $n_x$-th Fourier component of the velocity field with streamwise wavenumber, $ k_x= n_x \alpha$,  at the position $(y_i,z_i)$, 
 with $\dagger$ indicating the Hermitian transpose and $\alpha=2 \pi/L_x$ the smallest streamwise wavenumber in the channel.
The same point time-mean covariance is denoted $\C_{k_x} (y,z) $. From this 
time-mean covariance we obtain the time-mean Reynolds stresses
produced by the fluctuations.

 \section{R-S dynamical balance diagnostics}

The time-mean collocated R-S  and  the associated time-mean collocated fluctuation Reynolds stresses comprise 
components of the structure of the R-S and the dynamics maintaining it respectively.
We will now examine the
terms in  this equilibrium for the case of the low-speed streak.

 Equations \eqref{eq:NSm} and \eqref{eq:QLm} imply that the streamwise-mean streak, $U_s=U-[U]$, satisfies the equation
\begin{equation}
\partial_t U_s = - (V \partial_y (U)- [V \partial_y  U])- W \partial_z U_s -\partial_y (\overline{u v} - [\overline{u v}])- \partial_z (\overline{u w} - [\overline{u w}])+ R^{-1} \Delta U _s~,
\ee
so that, given that  $\partial_z [\overline{u w}]=0$, the time-mean  streak  satisfies  the force balance:
 \begin{equation}
 - \langle (V \partial_y (U)- [V \partial_y  U]) \rangle - \langle W \partial_z U_s \rangle -\partial_y( \langle \overline{u v} \rangle-\langle [\overline{u v}] \rangle )  - \partial_z \langle \overline{u w}  \rangle + R^{-1} \Delta \langle U _s \rangle = 0.
\label{eq:bal_U}
 \end{equation}
 The  terms comprising 
 this balance  are verified to be in a time-mean equilibrium in Fig.  \ref{fig:DNS_Us_balance}, for the NSE100,
 and in Fig.   \ref{fig:RNL_Us_balance},  for the RNL100. 
 %The  low-speed streak  in both NS100 and RNL100  is  in an essentially two-way balance. 
  Moreover, Fig. \ref{fig:DNS_Us_balance} and  Fig. \ref{fig:RNL_Us_balance} show that, in the time-mean,
the streak is principally supported 
 by the lift-up mechanism,
 $- \langle V \partial_y U -[V\partial_y U] \rangle$,
and opposed by  spanwise Reynolds stress divergence, 
$-\partial_z \langle \overline{u w} \rangle$ and diffusion $R^{-1} \Delta \langle U _s \rangle$.
% (cf. Fig. \ref{fig:DNS_Us_balance}d and  Fig. \ref{fig:RNL_Us_balance}d).
% and by dissipation, $R^{-1} \Delta \langle U _s \rangle$
%  (cf. Fig. \ref{fig:DNS_Us_balance}e and  Fig. \ref{fig:RNL_Us_balance}e).

 A typical time series of the low-speed streak at the centerline of the streak,  of the instantaneous  average streak acceleration by  the lift-up process, 
  $-\int_0^1 dy~ (V \partial_y U - [V \partial_y U])$, the average  acceleration 
by spanwise Reynolds stress divergence, $-\int_0^1 dy~ \partial_z \overline{uw} $,   and
the average acceleration  due to diffusion, $R^{-1} \int_0^1~dy~\Delta U_s$,  are shown in Fig. \ref{fig:streak_prod}.
 The time-mean acceleration and standard deviation over the entire dataset due to lift up is $-0.01~U_c^2/h$ (dashed blue) with $\sigma=0.004~U_c^2/h$, 
 that due to  Reynolds stress divergence
 is $0.007~U_c^2/h$ (dashed black) with $\sigma=0.004~U_c^2/h$ and that due to diffusion is $0.003~U_c^2/h$ (dashed green) with  $\sigma=0.0013~U_c^2/h$.
 Over the entire dataset the acceleration  due to  lift up and that due to Reynolds stress divergence, $-\int_0^1 dy~ \partial_z \overline{uw} $, 
are strongly correlated  with  cross-correlation coefficient  $0.72$ at lag $1.2 ~h / U_c$ 
as shown in Fig. \ref{fig:streak_prod1}. 
 Also, over the entire dataset streak maxima lead the streak regulation term, $ -\int_0^hdy~ \partial_z (\overline{uw})/(h^2 U_c^2)$,
 by  $2 ~h/U_c$.
 Two bursting  events are seen in Fig. \ref{fig:streak_prod} associated with the streak maxima at  $3275~h/U_c$ and $3866 ~h / U_c$.
 These streak maxima are followed by maxima  of the streak regulation term $ -\int_0^hdy~ \partial_z (\overline{uw})/(h^2 U_c^2)$ 
 at $3282~h/U_c$ and $3879 ~h / U_c$ respectively.  
 The near balance and near synchronicity seen in Fig. \ref{fig:streak_prod1} indicates that  the maintenance and regulation of the streak amplitude
 is occurring at all times and that breakdown events  are not primarily responsible for either the maintenance or the  regulation of the streak
 (see also Fig. \ref{fig:roll_prod}).

\begin{figure} 
 \vspace{1cm}
  \begin{center}
\includegraphics[width=1\textwidth]{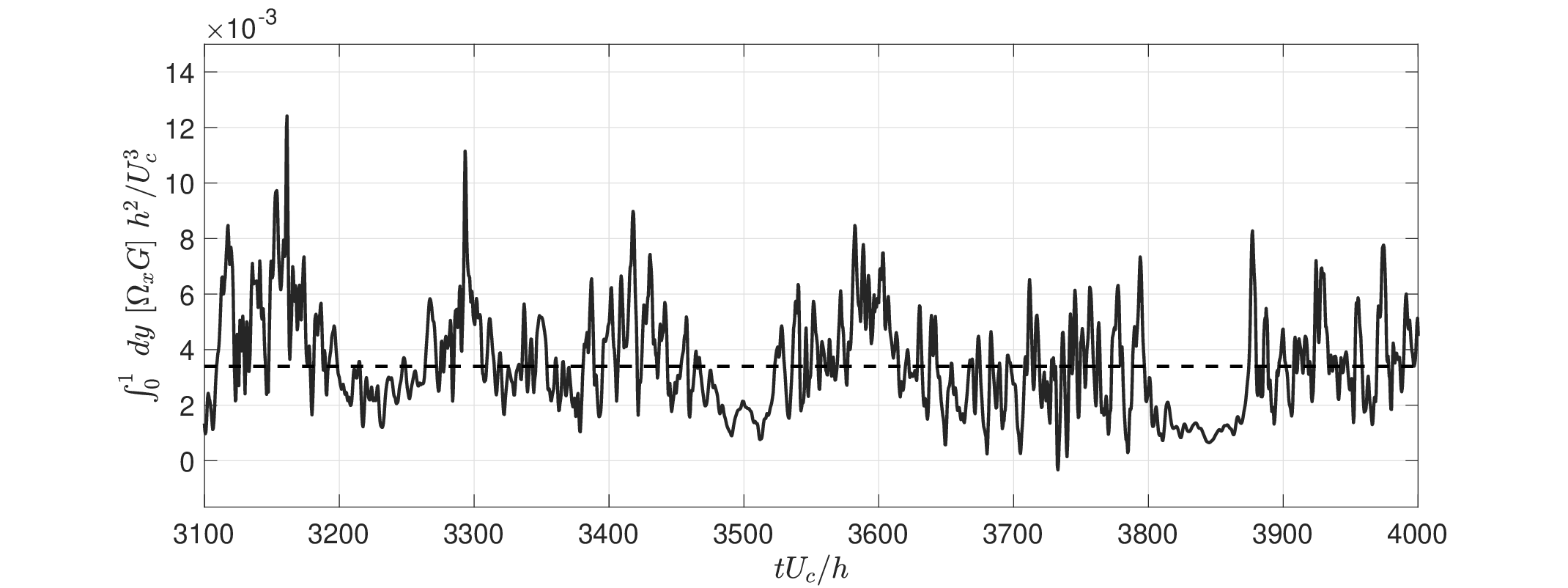} % figure 3a
%    \subfloat[]{\includegraphics[width=0.5\textwidth]{POD3d_inh/balance_stresses_rnl_Ox_full2_notuk.eps}\label{fig:RNL_Ox_balance}}  % figure 3b
 \end{center}
\caption{Typical section of the time series of the integrated correlation between the instantaneous value of the streamwise
mean vorticity and the streamwise mean vorticity source $G$, $ \int_0^1 dy ~\left [ \Omega_x G \right ]~h^2/U_c^3$,  for the case of the 
low-speed steak in NSE10. 
Over the whole dataset the time mean  is  $0.0035~h^2/U_c^3$ (dashed).
This figure shows that the forcing of the roll and consequently of the streak is continuous in time and almost always positive.}
\label{fig:roll_prod}
\end{figure}

 \section{Maintenance of the streamwise-mean roll}

Having verified the dominance of lift-up by roll circulations in supporting  the R-S,
our attention turns to study the mechanism giving rise to the remarkable 
universal coincidence in wall-turbulence of streaks with roll circulations properly configured to maintain them. By taking the curl of 
the streamwise-mean equations  \eqref{eq:NSm} and 
\eqref{eq:QLm}  we obtain that in both NSE100 and RNL10 the streamwise component of the vorticity  $\Omega_x = \partial_y W - \partial_z V$
 satisfies the equation:
\be
\partial_t \Omega_x =  \underbrace{-(V \partial_y + W \partial_z) \Omega_x}_A +\underbrace{(\partial_{zz} - \partial_{yy}) \overline{v w}+ \partial_{yz}(\overline{v^2} -\overline{w^2})}_{G} +  \underbrace{R^{-1} \Delta \Omega_x}_D~ , \label{eq:Omega}
\ee
in which the streamwise-mean wall-normal and spanwise  velocities are given by $V=-\partial_z \Delta^{-1} \Omega_x$ and $W=\partial_y \Delta^{-1} \Omega_x$
in which the inverse Laplacian $\Delta^{-1}$ incorporates the boundary conditions.
 
 \begin{figure} 
 \vspace{1cm}
  \begin{center}
\includegraphics[width=1\textwidth]{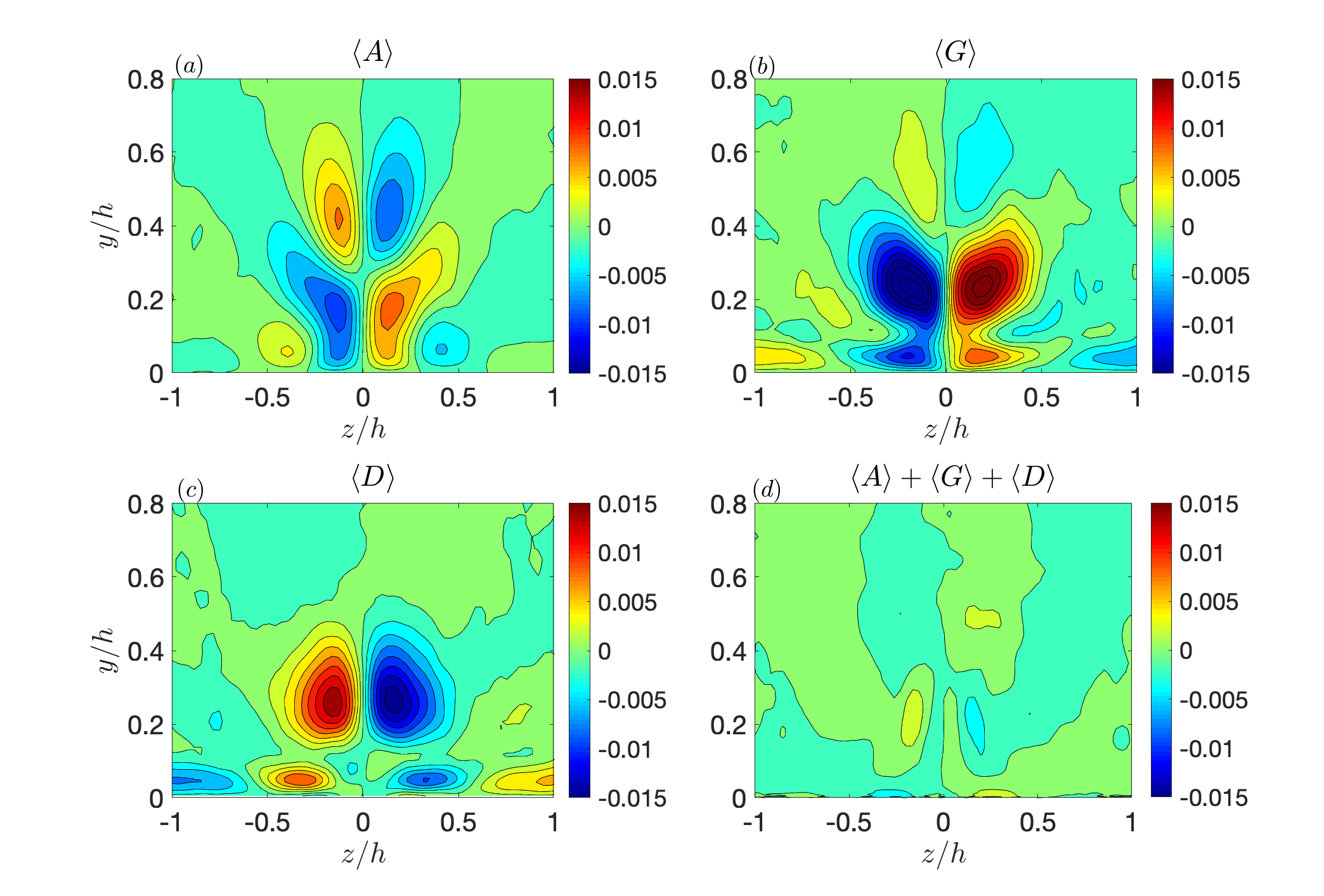} % figure 3a
%    \subfloat[]{\includegraphics[width=0.5\textwidth]{POD3d_inh/balance_stresses_rnl_Ox_full2_notuk.eps}\label{fig:RNL_Ox_balance}}  % figure 3b
 \end{center}
 \caption{For the low-speed streak in  NSE100  shown are contours in the $(y,z)$ plane of (a):  $\langle A \rangle=-\langle (V \partial_y + W \partial_z) \Omega_x \rangle$, 
 contribution to the time-mean
 rate of change of $\langle \Omega_x \rangle$ by roll self-advection, (b): $\langle G \rangle =(\partial_{zz} - \partial_{yy})  \langle \overline{v w} \rangle + 
 \partial_{yz} \langle (\overline{v^2} -\overline{w^2})\rangle$,  contribution to the time-mean  rate of change of $\langle \Omega_x \rangle$ by Reynolds stress divergence,
(c):   $\langle D \rangle=R^{-1} \Delta \langle \Omega_x \rangle$,  contribution to the time-mean  rate of change of $\langle \Omega_x \rangle$   by dissipation.
 The sum shown in (d) confirms that the above terms are in balance. The contour  interval is $0.0015~U_c^2/h$.}
  \label{fig:DNS_Ox_balance}  
\end{figure}

% - \langle V \partial_y (U_s + [U]) \rangle - \langle W \partial_z U_s \rangle -\partial_y \langle \overline{u v} \rangle - \partial_z \langle \overline{u w}  \rangle + R^{-1} \Delta \langle U _s \rangle = 0.
\begin{figure} 
 \vspace{1cm}
  \begin{center}
\includegraphics[width=1\textwidth]{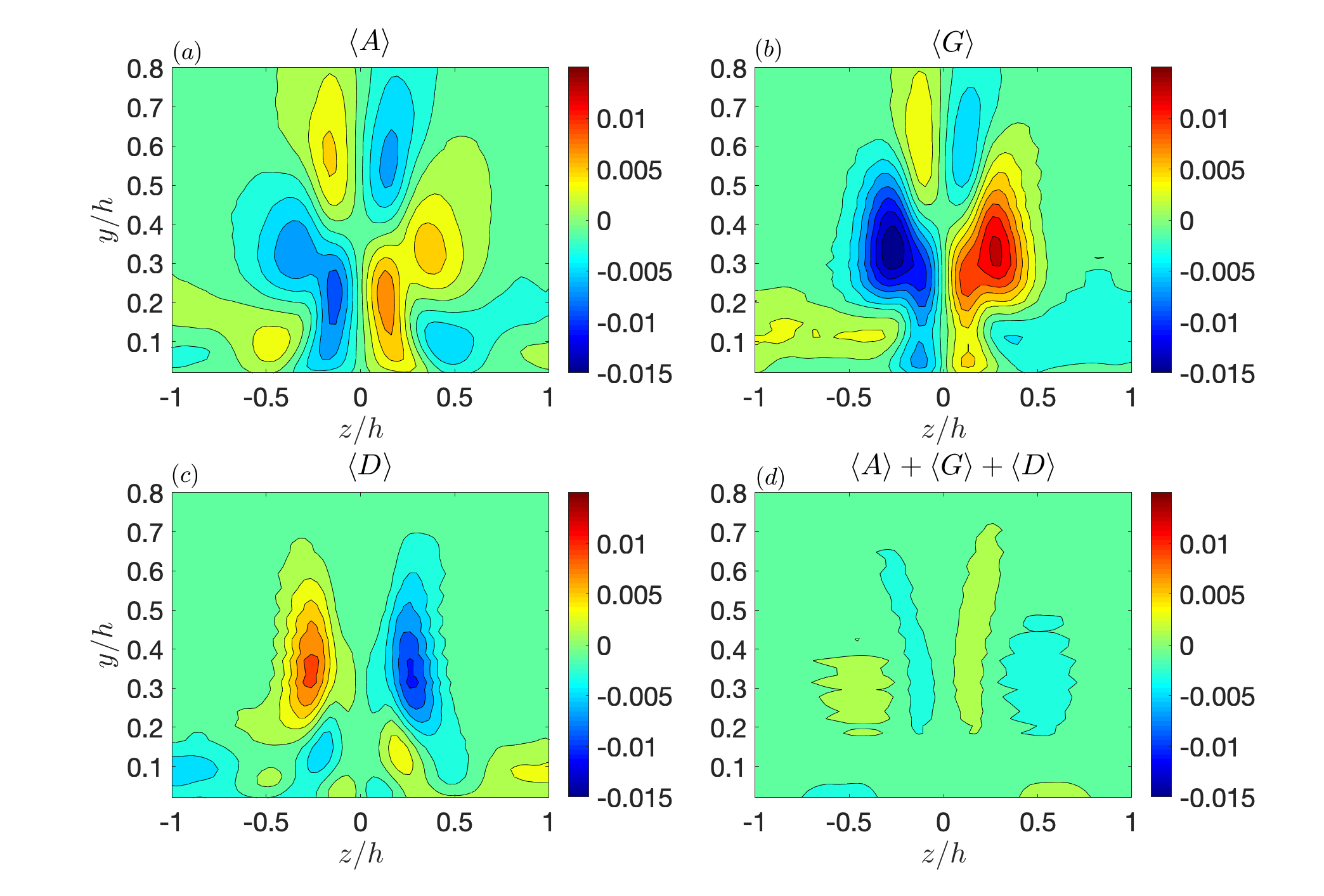} % figure 3a
%    \subfloat[]{\includegraphics[width=0.5\textwidth]{POD3d_inh/balance_stresses_rnl_Ox_full2_notuk.eps}\label{fig:RNL_Ox_balance}}  % figure 3b
 \end{center}
 \caption{As in Fig. \ref{fig:DNS_Ox_balance} for RNL100. The contour  interval is $0.0015~U_c^2/h$. }
 \label{fig:RNL_Ox_balance}
\end{figure}

The term $A$, representing advection of $\Omega_x$ by the roll velocities $(V,W)$, 
 is not a source of net streamwise vorticity. The roll vorticity is sustained against dissipation, $D$, 
 by    the curl of  the force arising from the Reynolds stress divergence, $G$. In this equation the
wall-normal component of the Reynolds stress divergence  force is:
\begin{equation}
F_y= -\partial_z(  \overline{v w})- \partial_y ( \overline{v^2}),
\end{equation}
while the spanwise component is:
\begin{equation}
F_z= -\partial_y (  \overline{v w})- \partial_z (  \overline{w^2}),
\end{equation}
which results   in the contribution to the rate of change of streamwise-mean vorticity in the streamwise direction: 
\begin{equation}
G=\hat{\xv} \cdot \nabla \times (0,F_y,F_z) = \left ( \partial_{zz} -\partial_{yy}  \right )  \overline{vw}+
\partial_{yz}  \left ( \overline{v^2} - \overline{w^2} \right ),
\label{eq:trq}
\end{equation}
where $\hat{\xv}$ is the unit vector in the streamwise direction.
The first RHS term in \eqref{eq:trq} represents the contribution to $G$
from the  Reynolds shear stress  $ \overline{vw}$,
while the second term represents the contribution  from  $  \overline{v^2} - \overline{w^2} $, which  can be 
identified with   anisotropy in  the Reynolds normal stress components.
The implications of this decomposition are discussed  by \cite{Alizard-etal-2021} in the context of 
the formation of streamwise constant rolls during transition to turbulence in the RNL framework.
% and by 
%\cite{Brundrett-1964,Perkins-1970,Gessner-1973} for diagnosing the  maintenance of persistent streamwise roll circulations in corner flows.
The Reynolds normal stress component of $G$  will be shown in the next section to dominate and determine the direction 
and location of the roll circulation and consequently of the streak acceleration.
% (cf. Fig. \ref{fig:DNS_Ox_balance}  and Fig. \ref{fig:RNL_Ox_balance}). %This was also found to be the case for roll formation in transitional RNL flows by  \cite{Alizard-etal-2021},  for the case of rolls in 
%turbulent duct flows \citep{Brundrett-1964,Perkins-1970,Gessner-1973},
% as well as in vortex-wave theory  (cf. \cite{Hall-Sherwin-2010}).}

% Fig. \ref{fig:DNS_Ox_balance}  and Fig. \ref{fig:RNL_Ox_balance} 
% verify that the vorticity is maintained by a two way balance between $G$ and $D$ given that
% advection is conservative. 
% The vorticity source $G$ is  collocated with the time-mean roll-circulation
% (cf. Fig. \ref{fig:ali2}a) 
%so that the induced streamwise mean wall-normal velocity  reinforces the low speed streak.
% Moreover,  $G$  reinforces the pre-existing streamwise-mean vorticity, $\Omega_x$,  as shown in
%  which shows 
 A time series of  the inner product of  $\Omega_x$ with $G$ is shown in Fig. \ref{fig:roll_prod}. 
 Two observations are appropriate: the first is  that  forcing by  Reynolds stresses is continuous in time and
almost always positive, the second is that streamwise-mean vorticity forcing  is negatively associated with bursting events such as that  occurring around $t=3800$. 
 %This figure shows that the reinforcement of the roll and consequently of the streak is continuous in time. 
 Continual  generation of streamwise vorticity supporting the existing roll circulation both in the buffer layer
and also in the logarithmic layer was previously documented
in  RNL turbulence at  Reynolds number $R_\tau=1000$  (cf. \cite{Farrell-etal-2016-VLSM}).  This result has not yet been confirmed  in DNS, but we expect it to be,
given that parallel mechanisms underlie wall-turbulence in RNL and DNS.
 
 \begin{figure} 
 \vspace{1cm}
  \begin{center}
    \includegraphics[width=1.0\textwidth]{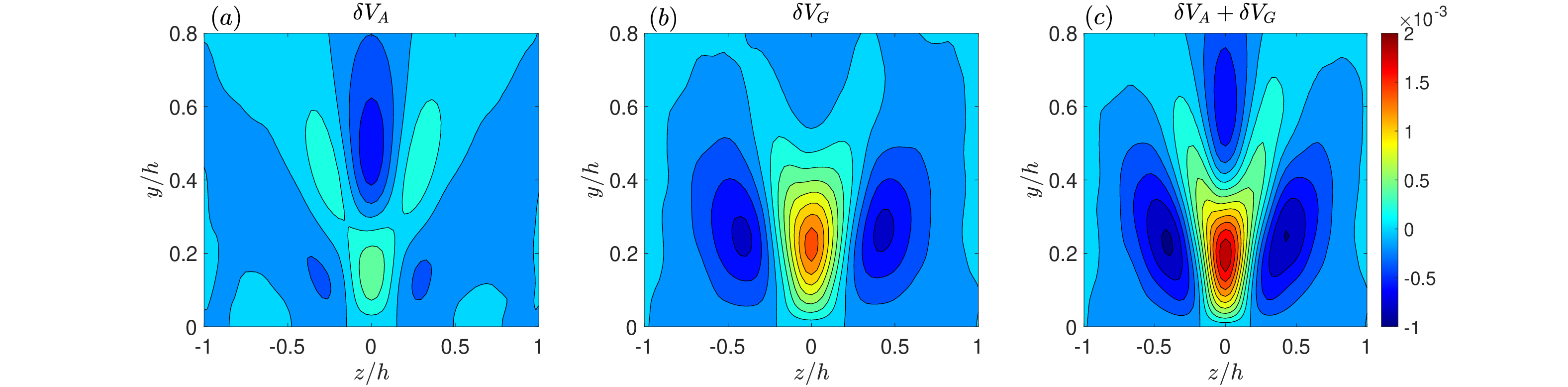}%
 \end{center}
 \caption{For the low-speed streak of NSE100 shown are: (a) the time-mean wall-normal velocity increment $\delta V_G$ 
 resulting from the Reynolds stress, $\langle G \rangle$, (b)
 the wall-normal velocity increment $\delta V_A$ from advection, $\langle A \rangle$,  and (c) their sum $\delta V_A+ \delta V_G$. 
 %In an infinite channel  the velocity vector field $(W,V)$ would be tangent to contour levels of $\Omega_x$ and the $\delta V_A$ would not
 %contribute to the roll maintenance. 
% The non-zero value of $\delta V_A$ reflects the deflection of the vector field 
% $(W,V)$ from tangency to  the contour levels of $\Omega_x$ induced by  pressure forces  in order
% to accommodate  the circulation instigated by $G$
%in the confines of the finite channel. 
This figure shows the contribution of the advection and Reynolds stress to the maintenance of the low-speed R-S.  
The associated $\langle A \rangle $, $\langle G \rangle $ fields are shown in Fig. \ref{fig:DNS_Ox_balance}a,b. The contour  interval is $2\times10^{-4}~U_c$. } 
\label{fig:deltaV_AG}
\end{figure}

 In the time-mean the  streamwise-mean vorticity, $\Omega_x$,   satisfies  the  balance:
  \begin{equation}
  \underbrace{- \langle (V \partial_y + W \partial_z) \Omega_x \rangle}_{\langle A \rangle}+
 \underbrace{(\partial_{zz} - \partial_{yy})  \langle \overline{v w} \rangle + \partial_{yz} \langle (\overline{v^2} -\overline{w^2})\rangle}_{\langle G \rangle} + \underbrace{R^{-1} \Delta \langle \Omega_x \rangle}_{\langle D \rangle} = 0.
 \label{eq:bal_Ox}  
 \end{equation} 
The three components of this time-mean  balance in  NSE100 and RNL100 are shown in 
 Fig. \ref{fig:DNS_Ox_balance}  and Fig. \ref{fig:RNL_Ox_balance}.

 The  roll  circulation resulting from the forcing by $\langle G \rangle $ can be understood  by assessing 
  the wall-normal velocity induced by $\langle G \rangle$ together with its modification by $\langle A \rangle$. The modification given by $\langle A \rangle$
  results from a pressure field  required so that the circulation forced by $\langle G \rangle$ satisfies boundary conditions.  
We project \eqref{eq:bal_Ox} to streamwise-mean  wall-normal velocity
 by multiplying  \eqref{eq:bal_Ox} with $-\delta t ~\partial_z \Delta^{-1}$  for 
 a  chosen time interval $\delta t$ in order    to obtain:
\begin{equation}
\delta V_A + \delta V_G = - \delta t~R^{-1} \Delta \langle V \rangle~,
\label{eq:deltaV_AG}
\end{equation}
 where $\delta V_G = -\delta t~\partial_z \Delta^{-1} \langle G \rangle$ is the wall-normal velocity increment induced by $\langle G \rangle$ over time interval $\delta t$
 and  $\delta V_A = -\delta t ~\partial_z \Delta^{-1} \langle A \rangle$ is the corresponding  wall-normal velocity increment induced by $\langle A \rangle$.
It is the  $\delta V_G$  induced by $\langle G \rangle$ and corrected by $\delta V_A$  that determines the equilibrium $V$ field as indicated in the  balance equation
\eqref{eq:deltaV_AG}.
The wall-normal velocity increments $\delta V_G$ and  $\delta V_A$   maintaining the low-speed R-S in NSE100 are shown in Fig. \ref{fig:deltaV_AG}.
This figure shows that $\delta V_G$ is  providing lift-up in the streak core supporting the low-speed R-S
and also the  corrective $\delta V_A$, which is   about $1/3$  of the $\delta V_G$, is adding to the  support of the R-S provided by $\langle G \rangle$.

In  Fig. \ref{fig:deltaV_AG} we have taken the time interval for the development of $\delta V$ to be $\delta t=1 ~h/U_c$;
however, it is instructive to identify a  physically relevant value of this time scale which for the low-speed streak is
given by $T_d/\delta t \approx V_{max}/ \delta V_{max}=12$, where $ V_{max}$ is the maximum wall normal velocity at the streak centerline  (cf. Fig. \ref{fig:ali2}a)
and $\delta  V_{max}$ is the maximum wall-normal velocity increment over  unit time also evaluated at the centerline (cf. Fig. \ref{fig:deltaV_AG}c).
This time scale can be interpreted as a Rayleigh damping time scale for equilibration of the roll circulation being forced  by the Reynolds stresses.

 \section{Contribution to roll forcing by the sinuous ($\cal S$) and varicose ($\cal V$) fluctuations}

In the previous section we showed that the Reynolds stresses induce vorticity forcing  that continuously reinforces the pre-existing
streamwise-mean streamwise vorticity  so as to sustain the R-S. 
Key to   understanding this remarkable property is the   dynamics 
of the ${\cal S}$  and ${\cal V}$ fluctuations collocated with the mean streak. 
{  In this section we isolate the $\cal S$ and $\cal V$  components of the 
velocity fluctuations  collocated with the streak and 
show that the maintenance of the mean R-S  can be attributed 
to the Reynolds stresses due to the ${\cal S}$ and ${\cal V}$ components of velocity acting independently.
Although instantaneous snapshots of the flow field would reveal Reynolds stresses arising from interaction between the ${\cal S}$ and ${\cal V}$ field,
this interaction vanishes in the time-mean.
This is expected because the R-S is mirror-symmetric and the non-vanishing of the time-mean ${\cal S}$ and ${\cal V}$ covariance would 
result in  Reynolds stresses incompatible with the mirror symmetry of the time-mean R-S. 
This will be  verified below.}

In order to define the time-mean covariance of the ${\cal S}$ and ${\cal V}$ components of the fluctuation field
we form at each time-step of the simulation the ${\cal S}$ and ${\cal V}$ components of the velocity field:
 \begin{equation}
\u_{\cal S} (x,y,z,t) = \frac{\u - \u_{mirror}}{2}~,~\u_{\cal V} (x,y,z,t) = \frac{\u + \u_{mirror}}{2} ~,
\end{equation}
in which the  mirror symmetric fluctuation field about the plane $z=0$ is defined as
\begin{equation}
{\u}_{mirror}(x,y,z,t) \equiv \begin{pmatrix}
u(x,y,-z,t) \\
v(x,y,-z,t) \\
-w(x,y,-z,t)
\end{pmatrix}.
\end{equation}

{ The  time-mean spatial covariances of the ${\cal S}$ and ${\cal V}$ components of the fluctuation field  at streamwise wavenumber $k_x$ 
are  $\C_{{\cal S}, k_x}(y_1,z_1,y_2,z_2) \equiv\< {\u}_{{\cal S}, k_x}(y_1,z_1) {\u}_{{\cal S}, k_x}^{\dagger}(y_2,z_2) \>$
and $\C_{{\cal V}, k_x}(y_1,z_1,y_2,z_2) \equiv\< {\u}_{{\cal V}, k_x}(y_1,z_1) {\u}_{{\cal V}, k_x}^{\dagger}(y_2,z_2) \>$, 
where $\u_{{\cal S}/{\cal V},k_x}$ are the Fourier amplitudes of the ${\cal S}$ and ${\cal V}$ components of the fluctuation velocity field at $k_x$,
while the corresponding covariance of the total field is given by $\C_{k_x}(y_1,z_1,y_2,z_2) \equiv\< {\u}_{k_x}(y_1,z_1) {\u}_{k_x}^{\dagger}(y_2,z_2) \>$.
The asymptotic approach in time of the equality %It was verified within numerical precision that 
\be
\C_{k_x}(y_1,z_1,y_2,z_2)=\C_{{\cal S}, k_x}(y_1,z_1,y_2,z_2) +\C_{{\cal V}, k_x}(y_1,z_1,y_2,z_2),
\label{eq:covar}
\ee  
has been verified, implying that there is no time-mean correlation between the ${\cal S}$ and ${\cal V}$ fluctuations.
Consequently,
the  time-mean fluctuation Reynolds stresses, which are a linear  function of the covariances,   are the sum of the Reynolds stresses obtained from the respective
${\cal S}$ and ${\cal V}$ covariances. The fluctuation Reynolds stress can be further partitioned into a  sum
over  $k_x$. Using this partition   into ${\cal S}$ and ${\cal V}$ components at each wavenumber $k_x$, 
we can separate  the contribution of the ${\cal S}$ and ${\cal V}$ components at each $k_x$ to the mechanism sustaining the R-S.}

\begin{figure*}
\centering
\includegraphics[width=34pc]{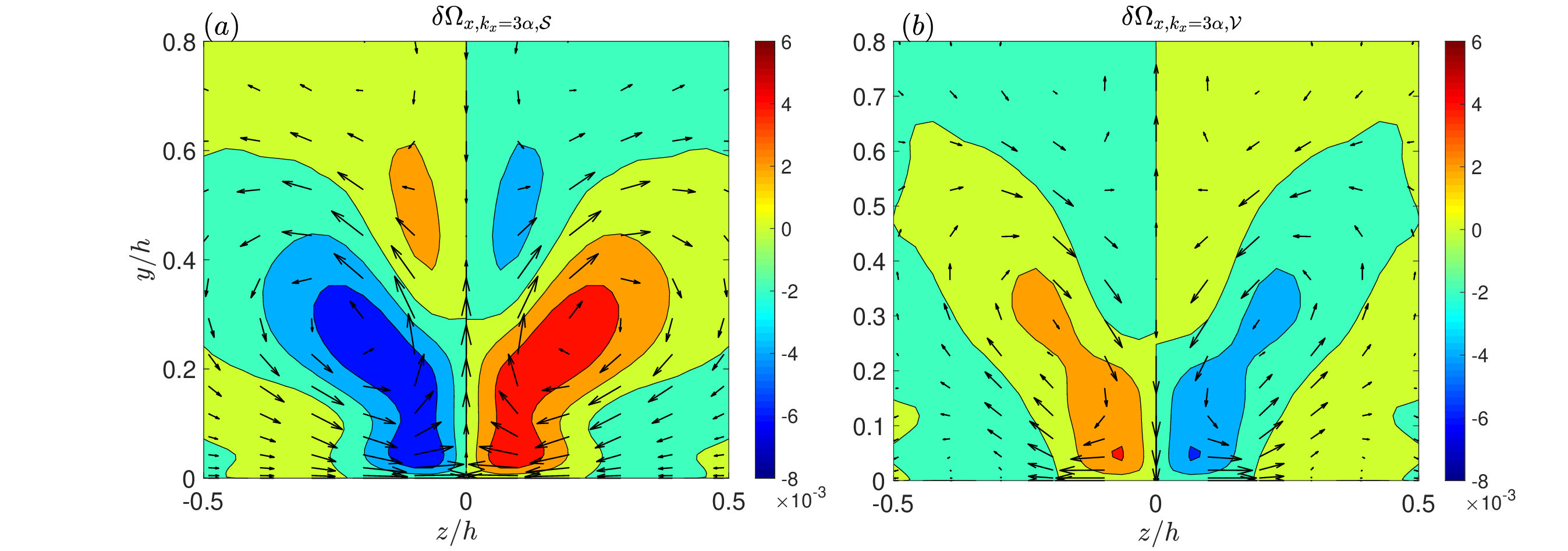}
%\subfloat[]{\includegraphics[width=16pc]{POD3d_inh/CPOD_stresses/lambda_cpod_y0p6_rnl_k2.eps}}
\vspace{-1em}
\caption{Increment of  mean streamwise vorticity, $\delta {\Omega}_{x,k_x}$,  induced over unit time by Reynolds stresses
of the ${\cal S}$ (panel a) and the ${\cal V}$ (panel b)   $k_x/\alpha=3$ fluctuations in  the mean low-speed streak in NSE100.
Wavenumber $k_x/\alpha=3$ is chosen because the forcing is maximized at this wavenumber (cf. Fig. \ref{fig:Veq_dns_rnl_kx}a). Also shown are 
vectors with components the   roll velocities induced over unit time $(\delta W_{k_x}, \delta V_{k_x})$. 
This figure shows that the ${\cal S}$ fluctuations reinforce the low-speed streak while the $\cal V$ fluctuations oppose it. Overall the ${\cal S}$ fluctuations
are dominant and the low-speed streak is sustained.   } \label{fig:dns_torque_omx}
\end{figure*}
 
We turn
now to study how the roll  is induced by the time-mean fluctuation Reynolds stresses.
We first consider the  contribution to the  roll forcing  by the time-mean Reynolds stresses due to  $k_x$ fluctuations, 
$\langle G_{k_x} \rangle=(\partial_{zz} - \partial_{yy})  \langle \overline{v w} \rangle_{k_x} + \partial_{yz} \langle (\overline{v^2} -\overline{w^2})\rangle_{k_x}$. 
This $\langle G_{k_x} \rangle$ acting alone would result, as discussed in the previous section,  in a wall-normal velocity increment over unit time,
$\delta V_{k_x} = -\delta t ~\partial_z \Delta^{-1} \langle G_{k_x} \rangle$,  
and a spanwise velocity increment over unit-time, $\delta W_{k_x} =\delta t ~\partial_y \Delta^{-1}  \langle G_{k_x} \rangle$.
The associated streamwise-mean vorticity increment is 
 $\delta \Omega_{x,k_x}=\langle G_{k_x} \rangle \delta t$  and $\Delta^{-1}$ the inverse Laplacian  required to account for the influence of pressure forces arising
from the boundary conditions. We choose $\delta t = 1$ from now on.

\begin{figure*}
\centering
\includegraphics[width=34pc]{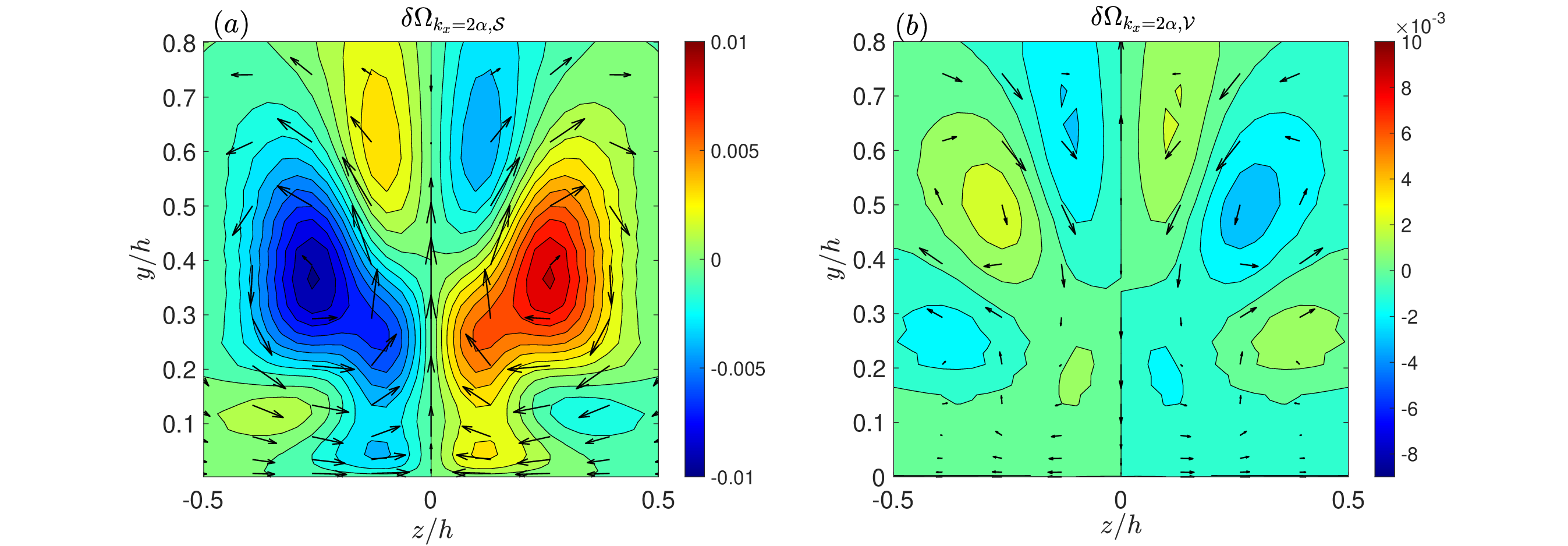}
%\subfloat[]{\includegraphics[width=16pc]{POD3d_inh/CPOD_stresses/lambda_cpod_y0p6_rnl_k2.eps}}
\vspace{-1em}
\caption{Increment of  mean streamwise vorticity, $\delta {\Omega}_{x,k_x}$,  induced over unit time by Reynolds stresses
of the ${\cal S}$ (panel a) and the ${\cal V}$ (panel b)   $k_x/\alpha=2$ fluctuations that are collocated with the mean low-speed streak in RNL100. 
Wavenumber $k_x/\alpha=2$ is chosen because the forcing is maximized at this wavenumber (cf. Fig. \ref{fig:Veq_rnl_kx}a).
Also shown are 
vectors with components the   roll velocities induced over unit time $(\delta W_{k_x}, \delta V_{k_x})$. 
This figure shows that the ${\cal S}$ fluctuations reinforce the low-speed streak while the ${\cal V}$ fluctuations oppose it as in  NSE100  shown in Fig. \ref{fig:dns_torque_omx}.    } \label{fig:rnl_torque_omx}
\end{figure*}

The spatial distribution
of $\delta \Omega_{x,k_x}$ and vector plots of the streamwise-mean velocity fields $(\delta V_{k_x}, \delta W_{k_x})$
induced  by ${\cal S}$ and ${\cal V}$ components of $k_x=3 \alpha$ fluctuations in NSE100 is shown  
in  Fig. \ref{fig:dns_torque_omx}. Note that the velocity increment vectors %while they are tangent to the contours of $\delta \Psi$ (not shown)
are not tangent to the contours of  the vorticity increments, $\delta \Omega_{x,k_x}$. This is due to the action of pressure forces arising due to the boundary conditions. 
This figure 
demonstrates that ${\cal S}$ Reynolds stresses produce mean vorticity that  reinforces
the low speed streak while  the ${\cal V}$ Reynolds stresses oppose the low speed streak.  In low-speed streaks the ${\cal S}$
Reynolds stresses dominate,
consistent with the ${\cal S}$ structures maintaining the low-speed streak.
 While both ${\cal S}$ and ${\cal V}$ fluctuations are present in association with low-speed streaks  
so that application of targeted data analysis techniques could be used to educe 
the presence of  e.g.  hairpin vortex structures  in 
association with low-speed streaks, this result demonstrates that the
varicose component at $k_x= 3 \alpha$ opposes rather than maintains the low-speed streak.
We will verify that this is also the case at other $k_x$.
Conversely, in high-speed streaks the ${\cal V}$  Reynolds stresses  dominate,  
consistent with maintaining the high-speed streak. 
We will show that this is also a general property.  In RNL100 we obtain similar results (cf. Fig. \ref{fig:rnl_torque_omx}).

  \begin{figure} 
 \vspace{1cm}
 
   \begin{center}
    \subfloat{
    \includegraphics[width=0.5\textwidth]{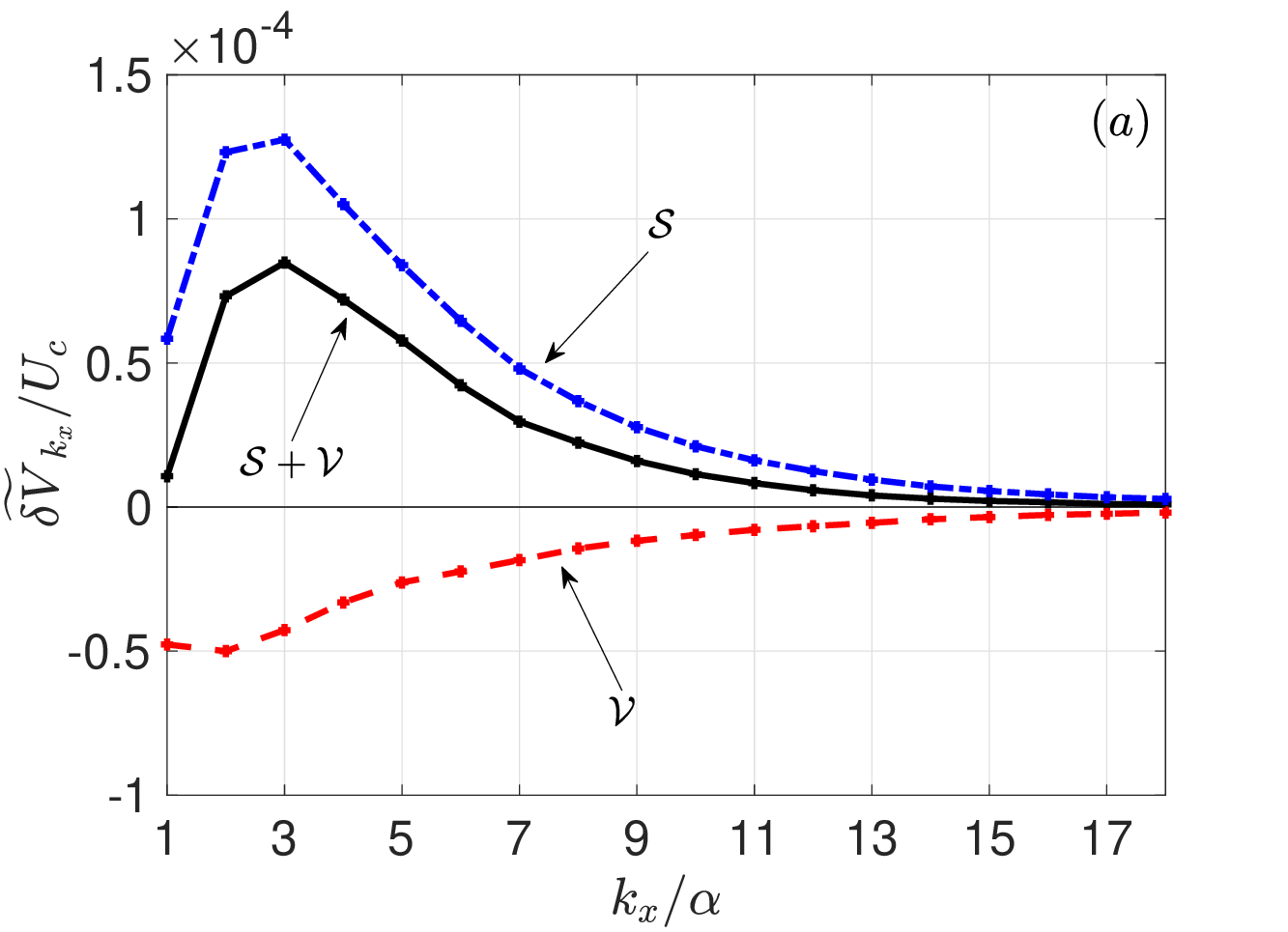}}%
      \subfloat{
       \includegraphics[width=0.5\textwidth]{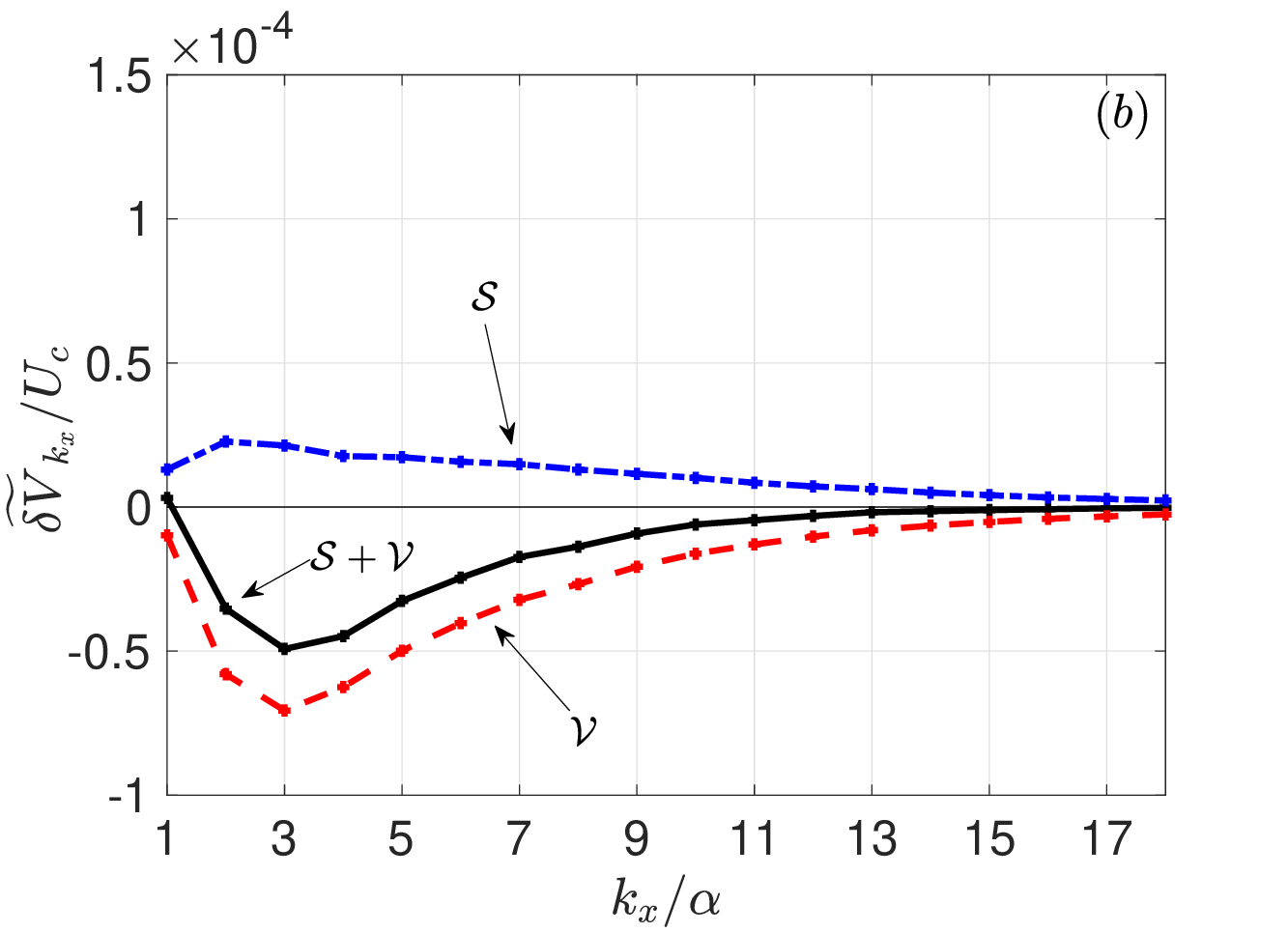}}%    \label{fig:dns_stresses_s_a}%} % figure 3a
          % \includegraphics[width=0.5\textwidth]{Vi_kx_sin_var_all_avg_rnl.eps}}%    \label{fig:dns_stresses_s_a}%} % figure 3a
 %   \subfloat[]{\includegraphics[width=0.5\textwidth]{POD3d_inh/POD_stresses/psi_psidot_rnl_k2_vw_v2w2.eps}
  %  \label{fig:rnl_var_tuk}}
     % figure 3b
 \end{center}
 \caption{ Velocity increments, $ \widetilde{\delta V}_{k_x}$,  forced by the Reynolds stresses over the primary area of lift-up,
  partitioned into ${\cal S}$ and  ${\cal V}$ components, 
 and the velocity increment induced by their sum, ${\cal S}$+${\cal V}$, as a function
 of the streamwise wavenumber of the fluctuations, $k_x/\alpha$, for the case of  the low-speed streak (panel (a)) and the high-speed streak (panel(b)) of NSE100.
 % The $\widetilde{\delta V}_{k_x}(y,z=0)$ at the centerline for each $k_x/\alpha$ is shown in Fig. \ref{fig:SVkx}.
 The largest induced velocity  occurs at $k_x/\alpha=3$ for both the low-speed streak  and  high-speed streak.
These figures show that in the time-mean the ${\cal S}$ fluctuations induce lift-up while the ${\cal V}$ induce push-down.
In the low-speed streak the $\cal S$ induced lift-up dominates the $\cal V$ push-down producing  maintenance of the low-speed streak,
while in the high-speed streak the $\cal V$  induced lift-up dominates the $\cal S$ push-down producing maintenance of the high-speed streak.}
%resulting in  the net  Reynolds stresses from the ${\cal S}$ and ${\cal V}$ fluctuations maintaining the inducing streak.}     
\label{fig:Veq_dns_rnl_kx}
\end{figure}

  \begin{figure} 
 \vspace{1cm}
 
  \begin{center}
    \subfloat{
    \includegraphics[width=0.5\textwidth]{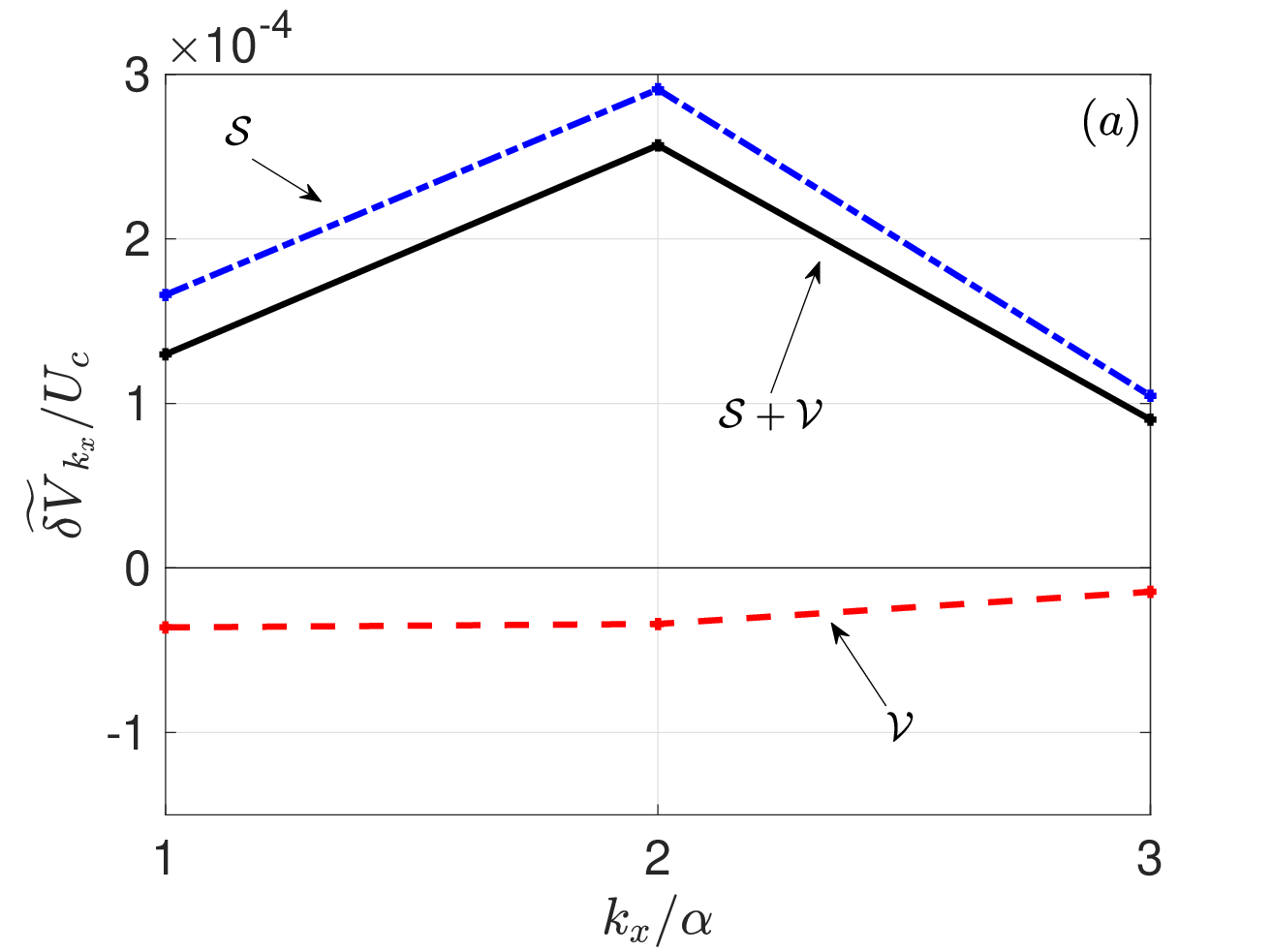}}%
      \subfloat{
       \includegraphics[width=0.5\textwidth]{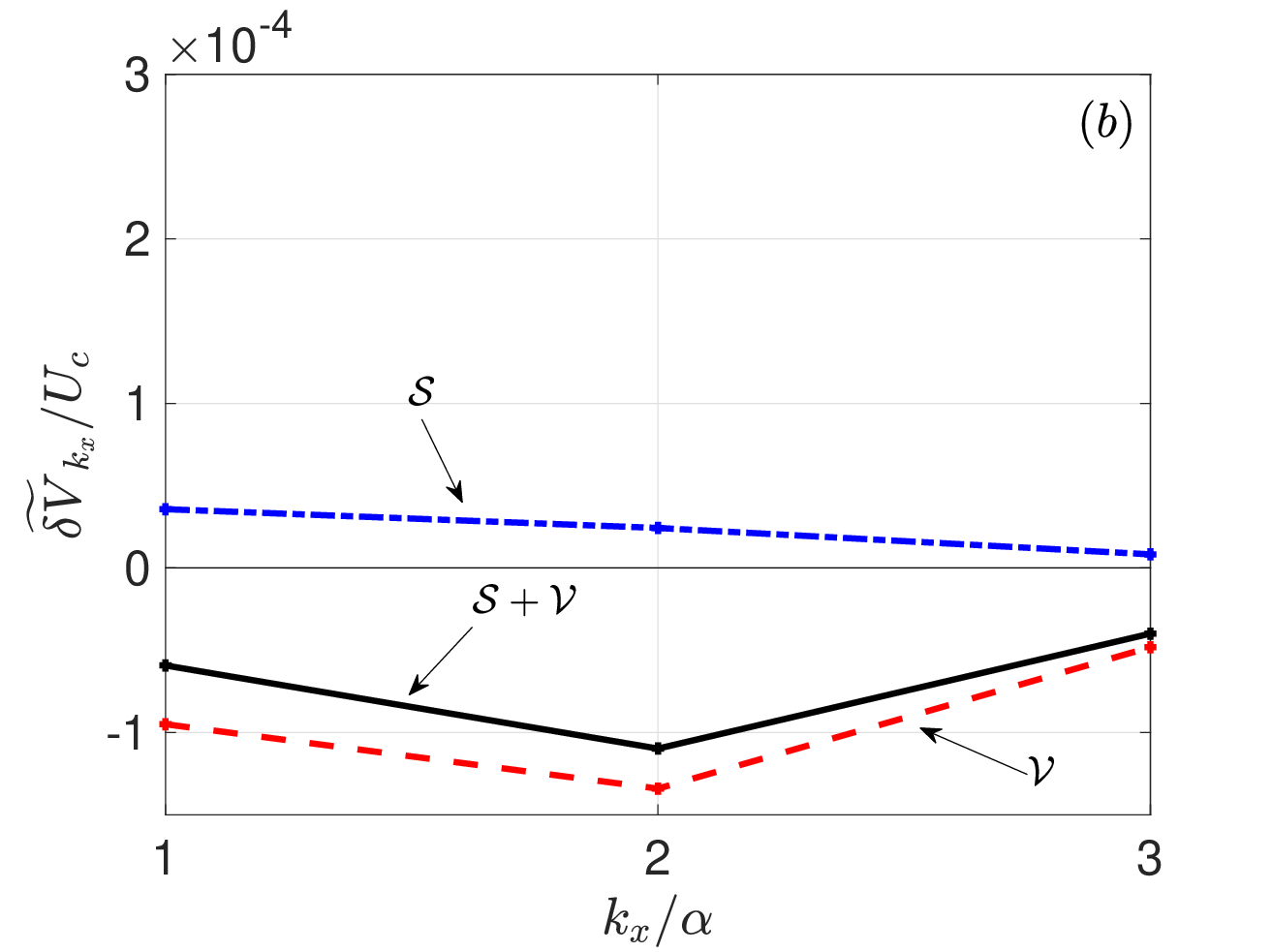}}%    \label{fig:dns_stresses_s_a}%} % figure 3a
          % \includegraphics[width=0.5\textwidth]{Vi_kx_sin_var_all_avg_rnl.eps}}%    \label{fig:dns_stresses_s_a}%} % figure 3a
 %   \subfloat[]{\includegraphics[width=0.5\textwidth]{POD3d_inh/POD_stresses/psi_psidot_rnl_k2_vw_v2w2.eps}
  %  \label{fig:rnl_var_tuk}}
     % figure 3b
 \end{center}
 \caption{  As in Fig. \ref{fig:Veq_dns_rnl_kx} except  RNL100.} 
\label{fig:Veq_rnl_kx}
\end{figure}

\begin{figure} 
 \vspace{1cm}
  \begin{center}
    \includegraphics[width=0.75\textwidth]{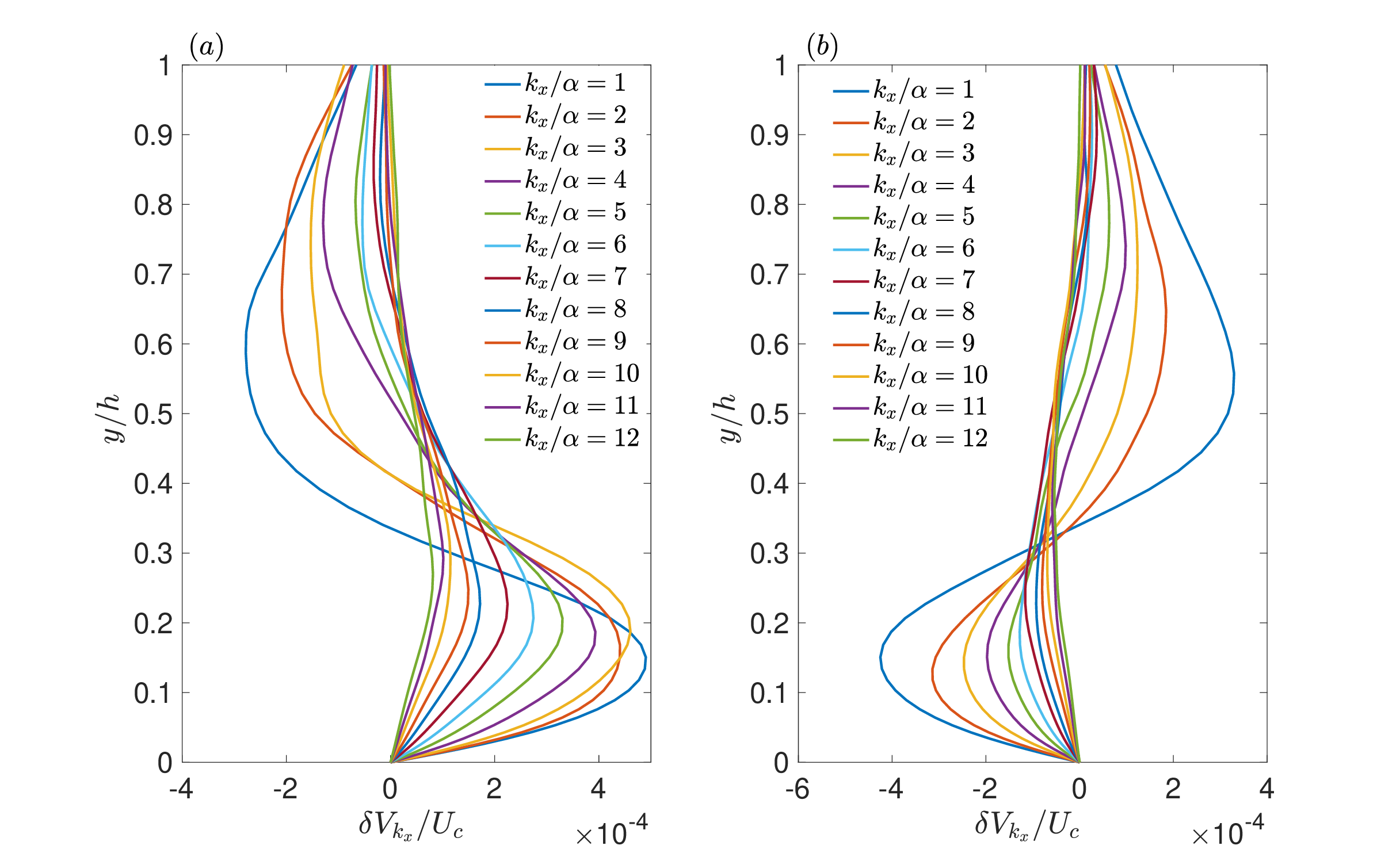}%
 \end{center}
 \caption{Wall-normal distribution  at the centerline  of  $ \delta V_{k_x}(y,z=0) $. 
 Shown separately are the ${\cal S}$ (a) and  the ${\cal V}$ (b) components   with  streamwise wavenumber $k_x/\alpha=1,2,\cdots,12$ for the case of  the low-speed streak  of NSE100.} 
\label{fig:SVkx}
\end{figure}

The velocity increment forced by the Reynolds stresses over the primary area of lift-up forcing (see Fig. \ref{fig:deltaV_AG}b) 
\begin{equation}
\widetilde{\delta V}_{k_x} \equiv  \int_{-z_0}^{z_0} \frac{dz}{2 z_0}  \int_0^{y_0} \frac{dy}{y_0}~\delta V_{k_x}(y,z),
\end{equation}
with $y_0=h/2$ and $z_0=0.26~h$, partitioned into ${\cal S}$ and  ${\cal V}$ components,  and the velocity increment induced by their sum, ${\cal S}$+${\cal V}$, as a function of the streamwise wavenumber of the fluctuations, $k_x/\alpha$, for the case of  the low-speed streak and the high-speed streak in NSE100
 is shown in Fig. \ref{fig:Veq_dns_rnl_kx}. 
The corresponding RNL100 results  shown in Fig. \ref{fig:Veq_rnl_kx}   are similar  to those of the NSE100, except
that in RNL100 the streak is supported by only the first three  streamwise wavenumbers, which are the streamwise wavenumbers
spontaneously retained by RNL dynamics.
These figures show that  in the time-mean and at all streamwise wavenumbers considered  the $\cal S$
fluctuations induce lift-up in both low and high-speed streaks, while the $\cal V$ induce  push-down.
These figures also show that in low-speed streaks the $\cal S$ component dominates the $\cal V$ at every $k_x$ 
resulting in the support of the low-speed streak,
while in high-speed streaks the $\cal V$ component  dominates the $\cal S$ resulting in the support of the high-speed streak. 
 Also the contribution to roll forcing by the fluctuations at each wavenumber is similarly distributed 
so that each wavenumber is contributing to the reinforcement of  the pre-existing R-S.
Note that the Reynolds stress induced time-mean $\widetilde{ \delta V}_{k_x}$ is maximized at $k_x/\alpha =3$ for both low-speed and high-speed streaks
in NSE100.
However,  the support of the streak extends over a broad band of  streamwise wavenumbers
implying structural robustness of the mechanism of roll forcing supporting the SSP cycle in
wall-turbulence.

Note that the
wall-normal velocity  increments induced either by the $\cal S$ or the $\cal V$ fluctuations in the case of the low speed streak are substantially
larger than the corresponding velocity {increments} induced for the case of the high-speed streak.  Moreover, 
the net roll forcing  from the sum of  the opposing ${\cal S}$ and $\cal V$  induced 
velocities  is approximately 2 times larger in the low-speed streak compared to the high-speed streak.
This dynamical advantage in forcing of the low speed streak in comparison to the forcing of 
the high speed streak, combined with the increased dissipation resulting 
from displacement of the high speed steak toward the boundary, provides 
explanation for the relative dominance of the low speed streak in observations of isolated R-S as  occur in this Poiseuille flow.
In the case of the highly-ordered R-S observed in wide channel Couette flow \citep{Pirozzoli-etal-2014}, the low and high-speed streaks would not be independent
and their mutual interaction would need to be taken into account.

 \section{Contribution to  streak forcing by the ${\cal S}$ and $\cal V$   shear and normal Reynolds stress components}
 
 Simplicity in analyzing the mechanism by which  the Reynolds stress forcing $G$ gives rise to the R-S can be obtained by
 concentrating on the forcing of $V$ along the centerline of the streak given by  $\delta V_{k_x}(y,z=0)$.
 The $y$ structure of the     $\cal S$ and $\cal V$ components of $\delta V_{k_x}(y,z=0)$  
 %and streamwise wavenumber of the contributing Reynolds stresses at the streak centerline 
 for  the low-speed streak in NSE100  is shown in Fig. \ref{fig:SVkx}. 
 This figure shows that  the Reynolds stress induced lift-up  
at each wavenumber  add  coherently.

 For the analysis of the streak forcing we 
choose to show  the streak velocity at the streak centerline  induced by $\delta V_{k_x}(y,0)$  
acting over  unit time, $\delta U = - \delta V_{k_x}(y,0) U'(y,0) \delta t $, with  $\delta t=1$ 
 and  $U'(y,0)$ 
 the shear of the streamwise flow at the streak centerline.   
 The streak velocity  $\delta U$ induced by the dominant $k_x/\alpha=3$ fluctuations is plotted in
 Fig. \ref{fig:DNS_low} for the low-speed streak in NSE100   and in Fig. \ref{fig:DNS_high} for the high-speed streak
 in NSE100,  both of which are located in the lower half of the channel by our collocation procedure. 
 The net $\delta U$ induced in the upper half of the channel by the $\cal S$ and $\cal V$ fluctuations,
 where there is no streak, vanishes in the time-mean. 
In the lower region, where there is  a streak,
the ${\cal S}$ and $\cal V$ contributions  do not cancel in the time-mean and  a 
net $\delta U$  results. In the low-speed streak region of  Fig.\ref{fig:DNS_low}a,
the $\cal S$ fluctuations dominate the $\cal V$ fluctuations in the time-mean  resulting in $\delta U$ increments  supporting the low-speed streak.
In general it can be shown that  $\cal S$ fluctuations force low-speed streaks while $\cal V$ fluctuations oppose this forcing \citep{Farrell-Ioannou-2022}.
In the high-speed streak regions, as shown in   Fig.\ref{fig:DNS_high}a, the high-speed streak is forced by 
the Reynolds stresses of $\cal V$ fluctuations which dominate the opposing tendency of the $\cal S$ fluctuations.  
The induced $\delta U$ in RNL100 are similar; for example, in
 Fig. \ref{fig:RNL_low} we show the induced $\delta U$ by the  $k_x/\alpha=2$ fluctuations in the low-speed streak in the lower-half channel of
 RNL100 and the $\delta U$ in the spanwise uniform flow in the upper-half channel. 
  
Partition of the  Reynolds stress induced streak increment $\delta U$ into the component  $\delta U_{vw}$ induced
 by the  Reynolds shear stresses,  $\langle \overline{vw} \rangle$,   
 and that induced  by the 
 Reynolds normal stresses, $\langle \overline{v^2-w^2} \rangle$ is shown  in Figures \ref{fig:DNS_low}b,c,    \ref{fig:DNS_high}b,c,   \ref{fig:RNL_low}b,c. 
 The  net Reynolds normal stresses, $\langle \overline{v^2-w^2} \rangle$,  which results from
 the dominance of the $\cal S$ over the $\cal V$ fluctuations in the presence of a low-speed streak, is seen to  determine the resulting net streak forcing.
  Moreover, similar distributions 
  characterize the induced acceleration for other streamwise-wavenumbers,  $k_x$. This will be shown below to be a  consequence of 
 universality in the structure of the Reynolds-stress with $k_x$.

  \begin{figure} 
 \vspace{1cm}
  \begin{center}
 \includegraphics[width=1\textwidth]{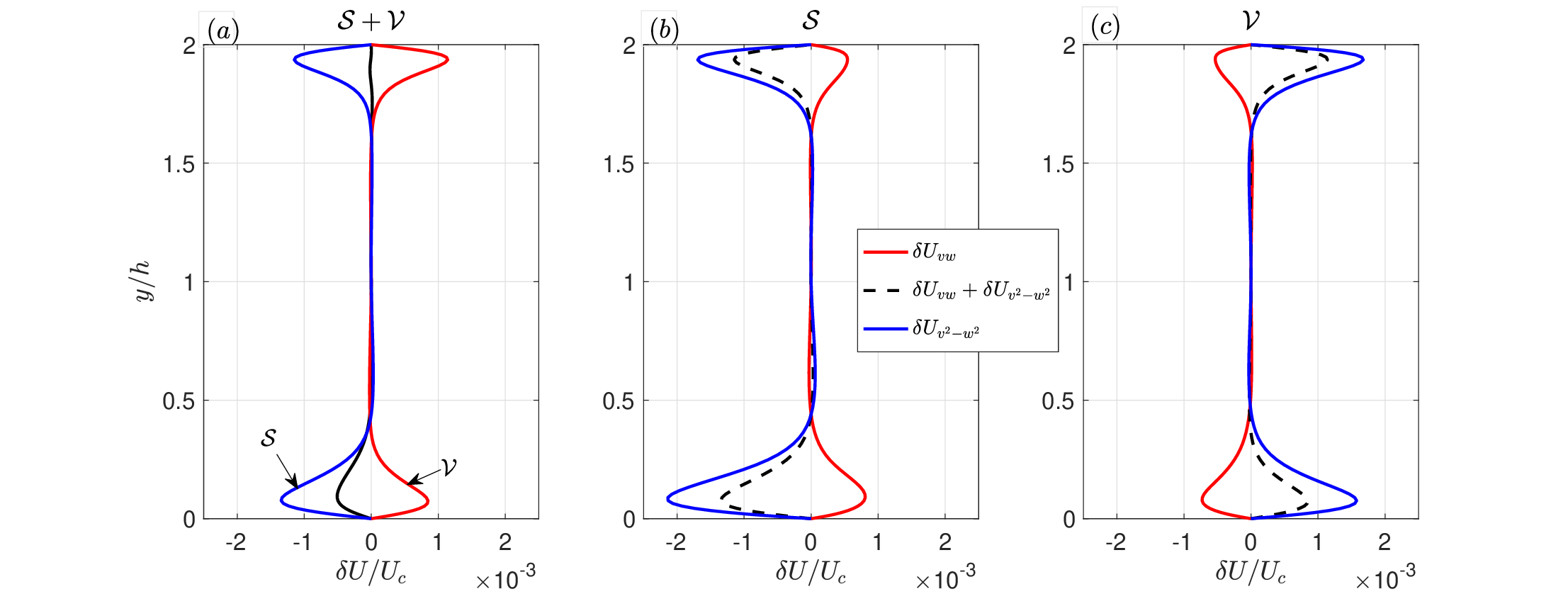} 
  %    \includegraphics[width=0.5\textwidth]{POD3d_inh/POD_stresses/Veq_dns_rnl.eps}% figure 3a
   % \subfloat[]{\includegraphics[width=0.5\textwidth]{POD3d_inh/POD_stresses/Vdot_Veq_rnl.eps}
     % figure 3b
 \end{center}
\caption{ \label{fig:dns_lift_s_a} In (a) is shown the contribution to streak forcing, $\delta  U$, 
that is induced by  lift-up. The lift-up is that induced over unit  time   by  the $k_x/\alpha=3$ fluctuations. Shown is the resulting $\delta U$
 in the low-speed streak region ($y/h<1$) and
in the spanwise uniform flow ($y/h>1$)  in NSE100 (black).  Shown separately are contributions to $\delta U$ induced by  the $\cal S$ (blue) and  $\cal V$ fluctuations  (red). 
%The net $\delta  U$  that results from their sum $\cal  S+V$  is in black. 
In 
(b)  is shown partition of the $\delta U$ induced by  $\cal S$  (dashed black) into the  component, $\delta U_{v^2-w^2}$,
 induced by  the $\langle \overline{ v^2-w^2} \rangle$ 
Reynolds stresses (solid blue) and  the component, $\delta U_{vw}$, induced by the $\langle \overline{vw} \rangle$  Reynolds stresses (solid red)
while in (c) is shown the corresponding partition for the $\cal V$ fluctuations. 
 This figure shows that $\cal S$ fluctuations tend to accelerate the low-speed streaks, while the $\cal V$ fluctuations
 tend to decelerate it, that the acceleration induced by the $\cal S$ is greater than that induced by the $\cal V$ in the 
 region of the low-speed streak in the lower half channel, that the ${\cal S}$ and $\cal V$ accelerations are equal and opposite
where there is no streak, and 
 that  the $\langle \overline{v^2-w^2} \rangle$ Reynolds normal stress dominates the forcing of lift-up resulting in streak forcing, $\delta U$.} 
 \label{fig:DNS_low}
\end{figure}

\begin{figure} 
 \vspace{1cm}
  \begin{center}
 \includegraphics[width=1\textwidth]{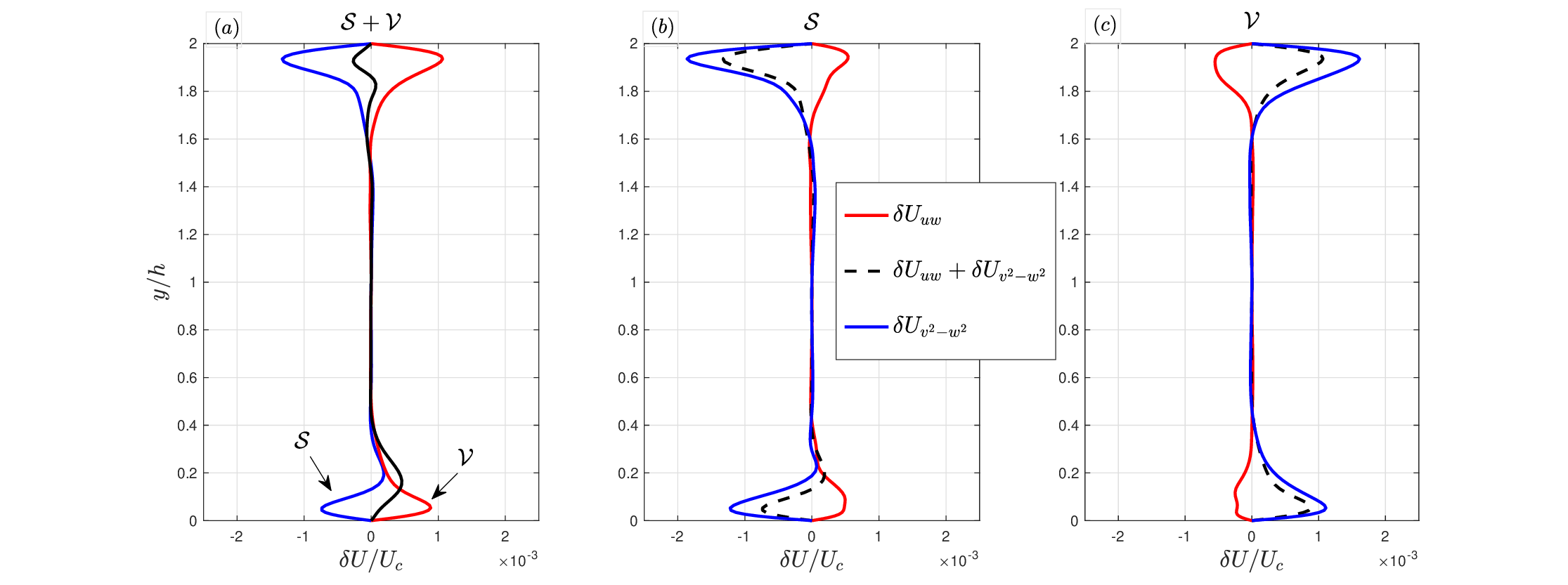} 
  %    \includegraphics[width=0.5\textwidth]{POD3d_inh/POD_stresses/Veq_dns_rnl.eps}% figure 3a
   % \subfloat[]{\includegraphics[width=0.5\textwidth]{POD3d_inh/POD_stresses/Vdot_Veq_rnl.eps}
     % figure 3b
 \end{center}
 \caption{As in Fig. \ref{fig:DNS_low}  except for  the $k_x/\alpha=3$ fluctuations in the high-speed streak in NSE100.}
 \label{fig:DNS_high}
\end{figure}
 
Indicative of the primary dynamics underlying the R-S  is the   Reynolds normal stress 
produced by the dominant wavenumbers. 
The distribution of the time-mean Reynolds normal stress  components partitioned into
the contribution from the $\cal S$ and the $\cal V$ fluctuations and the sum of these is shown in Fig. \ref{fig:SVstresses}.
The $\cal S$ fluctuations have $v=0$ and $\partial_z w=0$  at the centerline and   the normal stress  $\langle \overline{v^2-w^2} \rangle$ is 
negative and has a minimum as a function of $z$ at the centerline  with its overall minimum,    in the case of our streak,
attained  at  $y/h\approx 0.4 $ above   the center
of the streak, which is at $y/h \approx 0.15$ (cf. Fig. \ref{fig:ali2}a and Fig. \ref{fig:SVstresses}a). 
The $\cal V$ fluctuations have  $w=0$ and $\partial_z v =0$ at the centerline 
 consistent with  maxima of $\langle \overline{v^2-w^2} \rangle$ at the center of the streak (cf. Fig. \ref{fig:SVstresses}b).
In the absence of a streak, as in the region of the upper boundary of the channel,
the $\cal S$ fluctuations and the $\cal V$ fluctuations are equal and
the sum $\langle \overline{v^2-w^2}\rangle $ is constant in the spanwise direction, as shown in  Fig. \ref{fig:SVstresses}c near the upper boundary, 
and no roll forcing results
from the Reynolds normal stress.
In low-speed streaks, as in the region of the lower boundary of the channel,  the $\cal S$ fluctuations dominate consistent with the primacy of this term in providing
the required roll forcing
to maintain the low-speed
streak through lift-up (cf.  Fig. \ref{fig:SVstresses}c).
In contrast to the case of low-speed streaks, in high-speed streaks the $\cal V$ fluctuations dominate with 
the maximum  $\langle \overline{v^2-w^2}\rangle $ of the $\cal V$ fluctuations
almost canceling the minimum of the $\cal S$ fluctuations at the centerline
leading  the total $\langle \overline{v^2-w^2} \rangle$ to be determined by the two minima of
$\langle \overline{v^2-w^2} \rangle $ of the $\cal V$ fluctuations at the wings of the streak   (cf. Fig. \ref{fig:SVstresses_high}).
Note that the stresses in the presence of a high-speed streak are not mirror images of the stresses in the presence of a  low-speed streak
as the  low-speed streak flow is not a mirror image of the high-speed streak flow. However,
as  discussed in the next section, the stresses in the presence of infinitesimal low and high streaks are mirror images of each other
and remarkably the low speed streak is dominantly forced by the $\cal S$ fluctuations and the 
high-speed streak by $\cal V$ fluctuations even for infinitesimal strain of the perturbation field.
The high-speed streak  is supported by the $\langle \overline{w^2}\rangle$ component of the 
normal stress at the wings of the high-speed streak,
consistent with the $\langle \overline{w^2} \rangle$ Reynolds stress distribution being the dominant component supporting
both  low and high speed streaks. A similar dominance of the $\overline{w^2}$ component of the normal stress  in  roll formation was found 
in transitional RNL flows by  
\cite{Alizard-etal-2021}  and in the vortex-wave theory for the generation of rolls   (cf. \cite{Hall-Sherwin-2010}).
%Earlier experimental work  on
%turbulent duct flows by \cite{Brundrett-1964}  isolated  the $\langle \overline{v^2-w^2} \rangle$ as the component of the Reynolds stress  responsible for 
%sustaining the corner rolls but without distinguishing between the $\langle \overline{v^2} \rangle$ and $\langle \overline{w^2} \rangle$ contributions.}

 The crucial observation is that in the region of the streak an asymmetry between the ${\cal S}$ and $\cal V$ induces net Reynolds stresses 
 that sustain the pre-existing streak.  This asymmetry  between ${\cal S}$ and $\cal V$  fluctuations 
 arises as a general property of turbulence in the presence of a streak, as will be argued in the next section,  
 and manifests in the time-mean statistics as a general property
that is responsible for the roll forcing that generates and maintains the SSP and that underlies the universal 
mechanism of the S3T {modal} instability of spanwise uniform flows reponsible for the emergence of the R-S as a ubiquitous structure 
in  turbulent shear flows \citep{Farrell-Ioannou-2012,Farrell-Ioannou-2017-bifur,Farrell-Ioannou-2022}.%Farrell-Ioannou-2023-S3T}

\begin{figure} 
 \vspace{1cm}
  \begin{center}
 \includegraphics[width=1\textwidth]{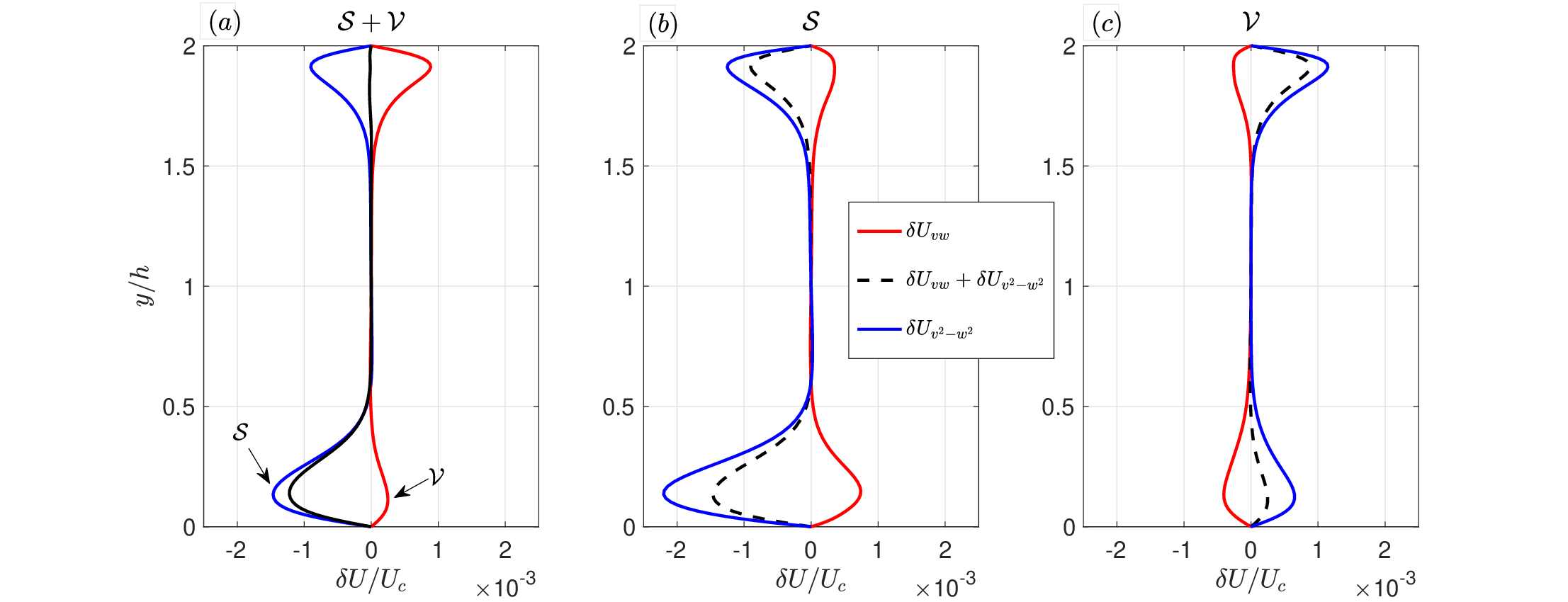} 
  %    \includegraphics[width=0.5\textwidth]{POD3d_inh/POD_stresses/Veq_dns_rnl.eps}% figure 3a
   % \subfloat[]{\includegraphics[width=0.5\textwidth]{POD3d_inh/POD_stresses/Vdot_Veq_rnl.eps}
     % figure 3b
 \end{center}
\caption{ As in Fig. \ref{fig:DNS_low}  except for the $k_x/\alpha=2$ fluctuations in the low-speed streak in RNL100.}
 \label{fig:RNL_low}
\end{figure}

\begin{figure} 
 \vspace{1cm}
  \begin{center}
 \includegraphics[width=1\textwidth]{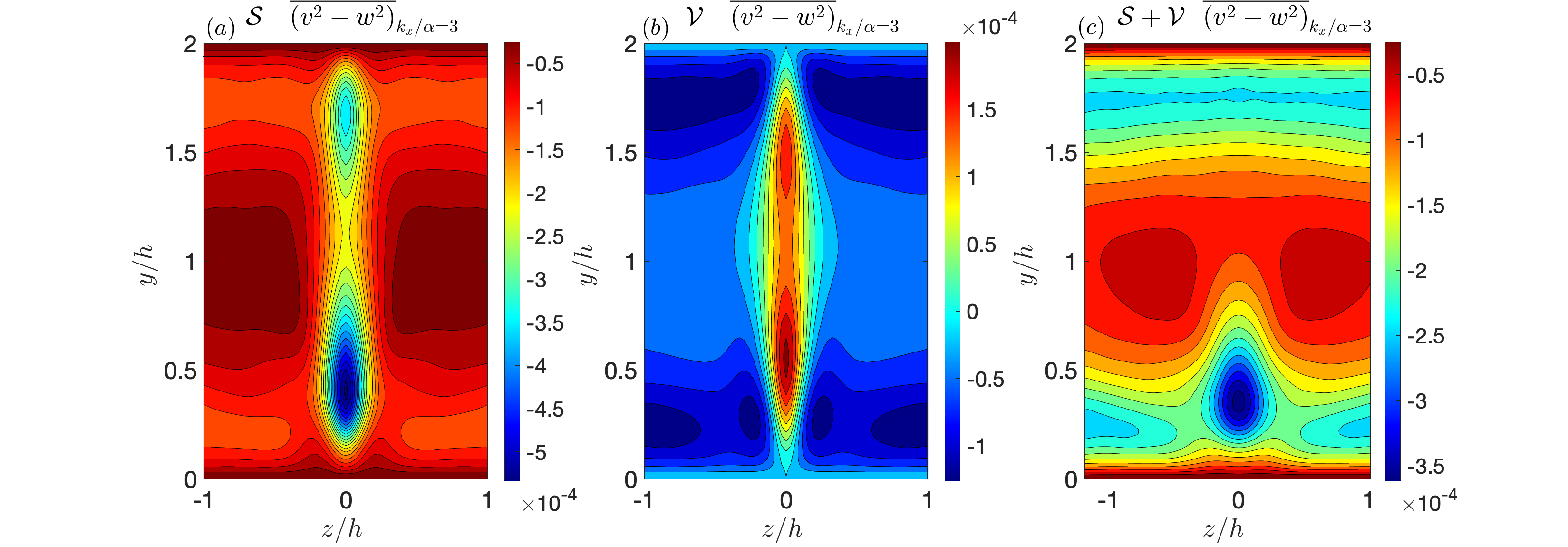} 
  %    \includegraphics[width=0.5\textwidth]{POD3d_inh/POD_stresses/Veq_dns_rnl.eps}% figure 3a
   % \subfloat[]{\includegraphics[width=0.5\textwidth]{POD3d_inh/POD_stresses/Vdot_Veq_rnl.eps}
     % figure 3b
 \end{center}
\caption{Time-mean Reynolds normal stress at $k_x/\alpha=3$  in NSE100  for the low-speed streak shown in Fig. \ref{fig:ali2}.
The normal stress shown is partitioned into $\cal S$  (panel(a))
and $\cal V$  (panel (b)) components. The total time-mean  normal stress is the sum of $\cal S+V$ (panel (c)).
This figure shows that  the low-speed streak results primarily from the $\cal S$  component.
Near the upper boundary  the flow is spanwise homogeneous and the normal stress becomes spanwise constant
producing no roll-forcing. The contour interval is $0.25\times 10^{-4}~U_c^2$. }
 \label{fig:SVstresses}
\end{figure}

 \begin{figure*}
\centering
\includegraphics[width=32pc]{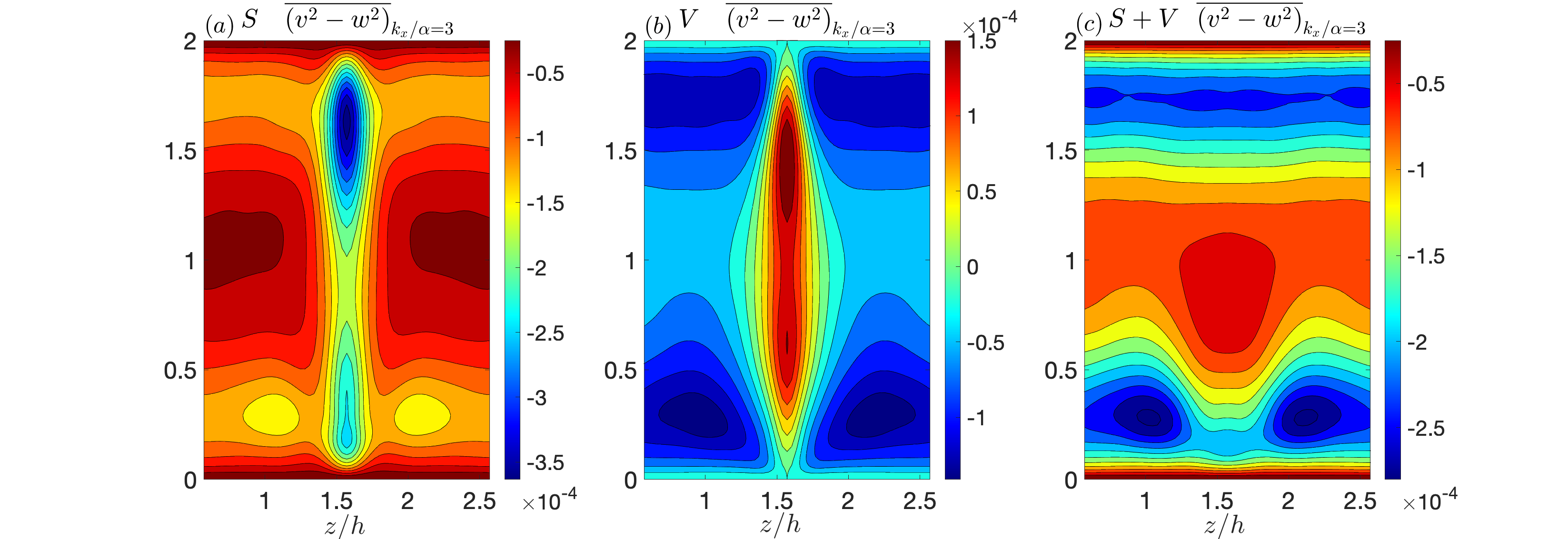}
\vspace{-1em}
\caption{As in Fig. \ref{fig:SVstresses} except for the high-speed streak.}
\label{fig:SVstresses_high}
\end{figure*}

 \begin{figure*}
\centering
\includegraphics[width=32pc]{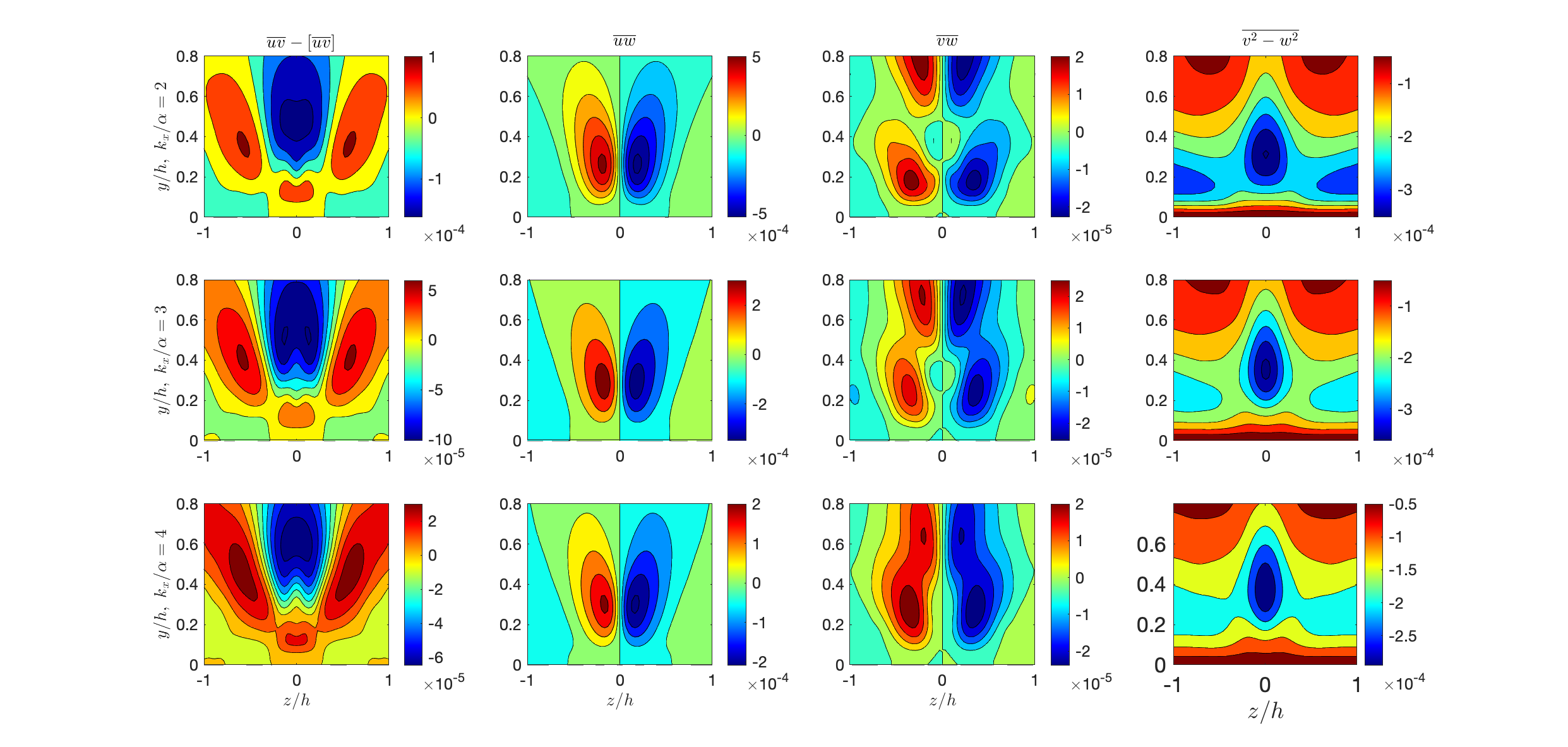}
\vspace{-1em}
\caption{Time-mean  Reynolds stresses of fluctuations 
collocated with
low-speed streak of NSE100. Top row:  $k_x/\alpha=2$; middle row  $k_x/\alpha=3$; bottom row $k_x/\alpha=4$.
The first two columns show contours of $\langle \overline{uv} - [\overline{uv}] \rangle$ and
$\langle \overline{uw}  \rangle$ which comprise the Reynolds stresses responsible for the regulation of the streak.
The third and fourth column show contours of $\langle \overline{vw} \rangle$ and
$\langle \overline{v^2-w^2}\rangle$  which comprise the Reynolds stresses responsible for forcing
the roll sustaining the low-speed streak.
 This figure shows that there is universality in the mechanism sustaining and regulating the low speed streak as a function of streamwise wavenumber.}
\label{fig:stresses_dns_low}
\end{figure*}

 \begin{figure*}
\centering
\includegraphics[width=32pc]{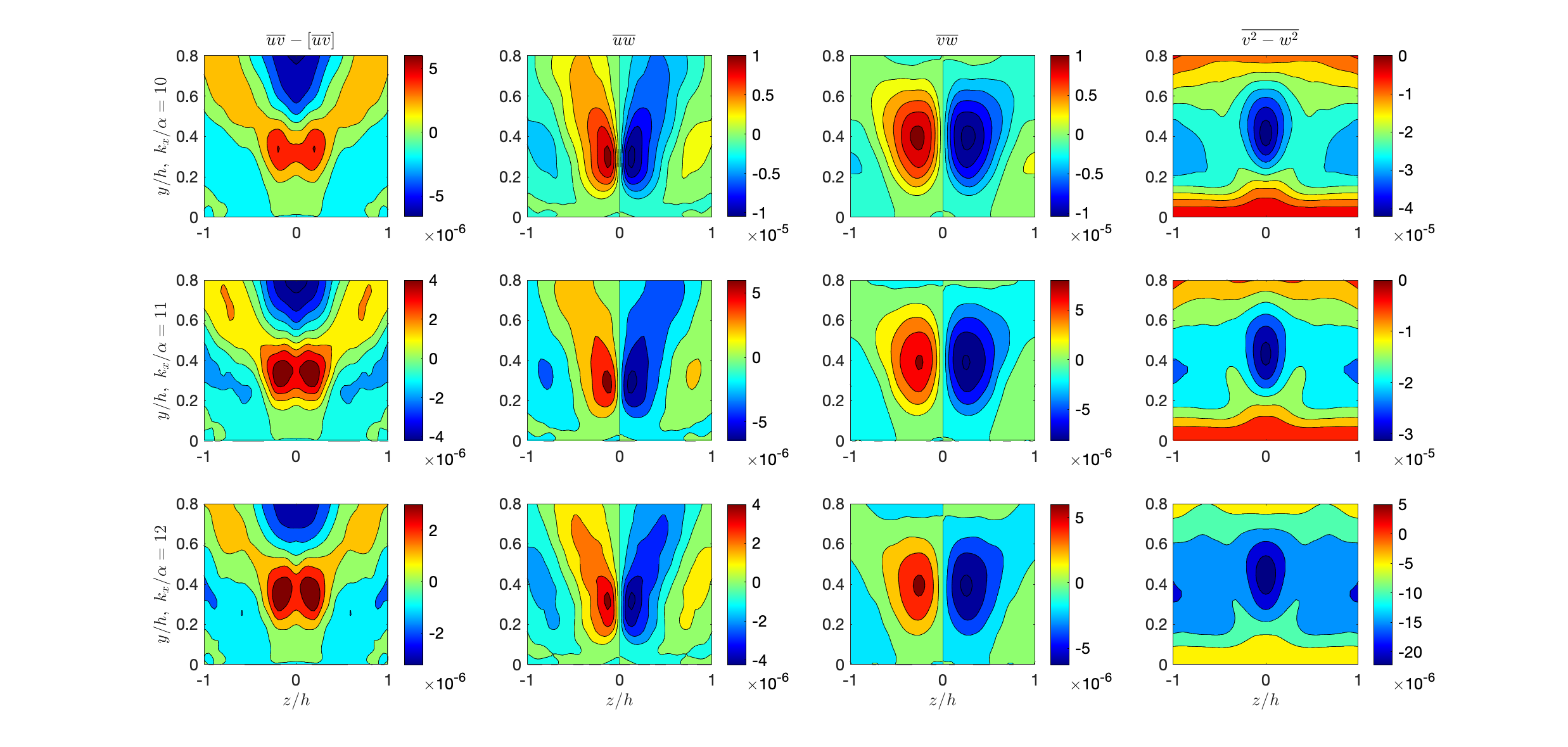}
\vspace{-1em}
\caption{ As in Figure \ref{fig:stresses_dns_low} 
for fluctuations with  $k_x/\alpha=10$ (top row), $k_x/\alpha=11$ (middle row)  
and $k_x/\alpha=12$ (bottom row).  This figure shows that the universality in structure is still apparent at high streamwise wavenumbers.}  \label{fig:stresses_10_12}
\end{figure*}

\begin{figure*}
\centering
\includegraphics[width=32pc]{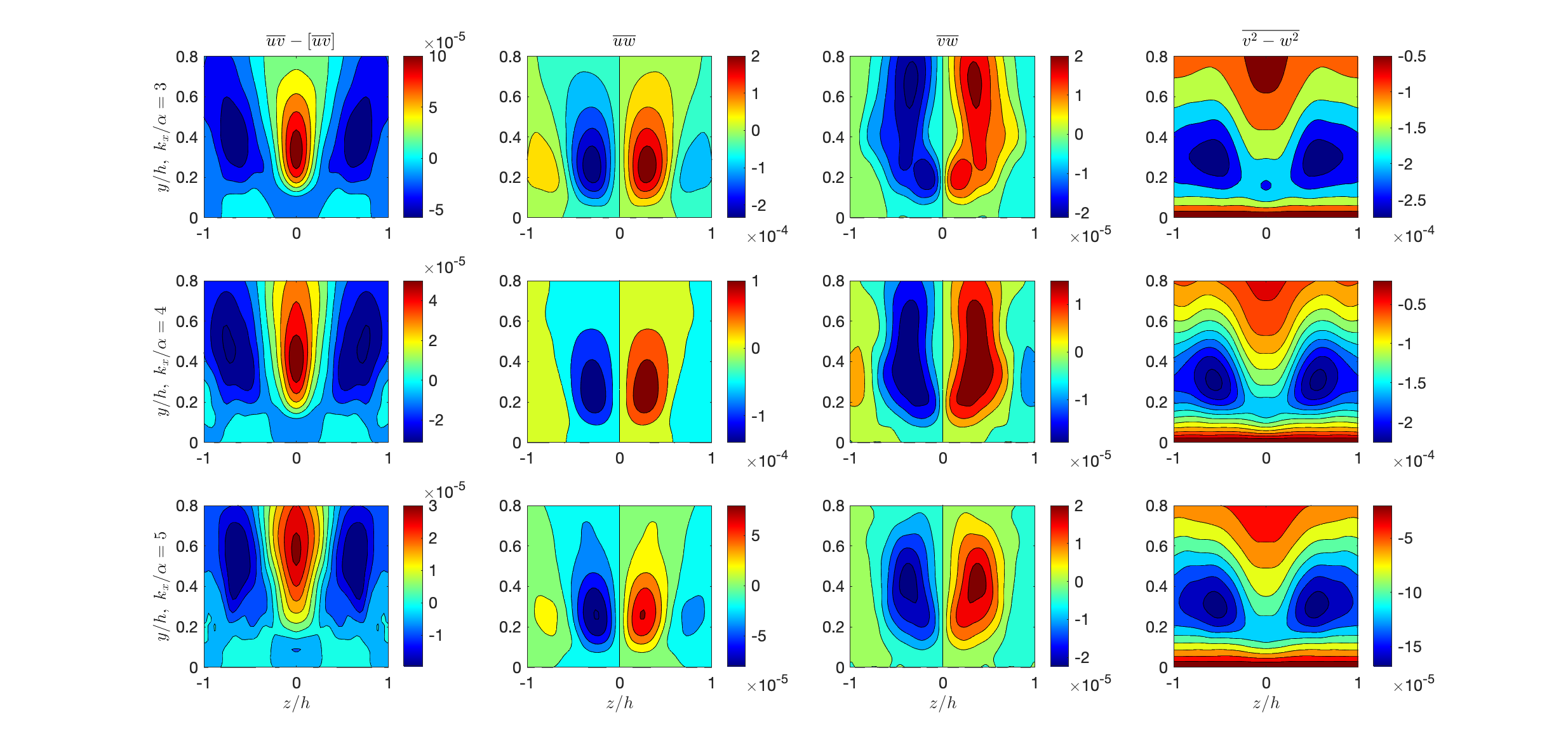}
\vspace{-1em}
\caption{As in Figure \ref{fig:stresses_dns_low} 
except for fluctuations with  $k_x/\alpha=3$ (top row), $k_x/\alpha=4$ (middle row)  
and $k_x/\alpha=5$ (bottom row) in the high-speed streak. } \label{fig:stresses_dns_high35}
\end{figure*}

  \begin{figure*}
\centering
\includegraphics[width=32pc]{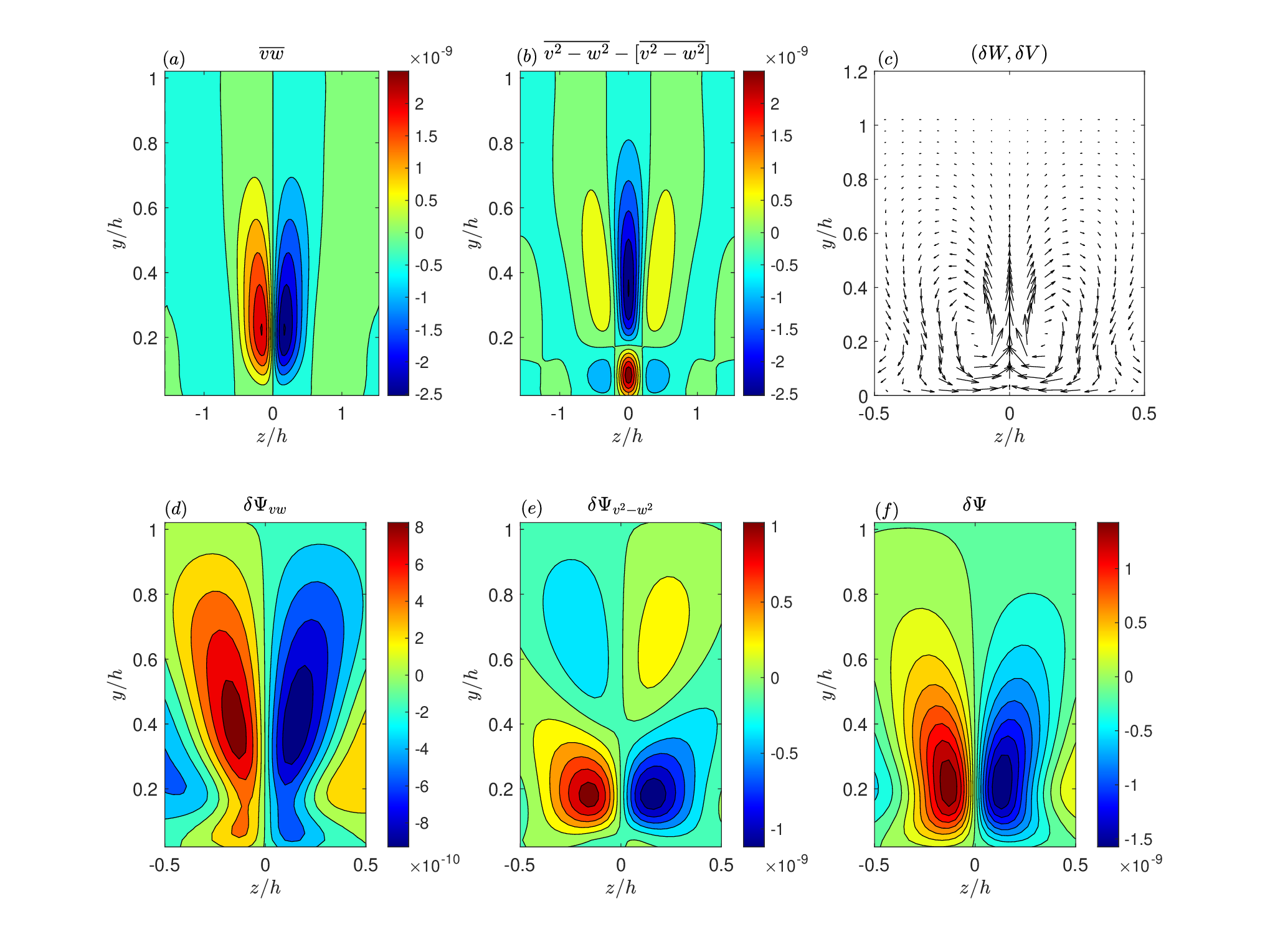}
\vspace{-1em}
\caption{Reynolds stresses predicted by  the STM about the NSE100  low-speed streak: the ensemble mean Reynolds shear 
stress  $\langle \overline{v w} \rangle$ (panel (a)), and the Reynolds normal stress component
$\langle \overline{ v^2- w^2}  \rangle$ (panel (b)).
Also shown in (c)  is the roll circulation $(\delta W, \delta V)$ 
 induced in unit time by both components of the Reynold stress. 
 The streamfunction  $\delta \Psi_{vw}$ of the roll circulation induced by $\langle \overline{v w} \rangle$ is shown in (d) and the streamfunction $\delta \Psi_{v^2-w^2}$ of the roll circulation induced by
$\langle \overline{ v^2- w^2}\rangle $ is shown in (e), while
the total streamfunction $\delta \Psi=  \delta \Psi_{vw}+\delta \Psi_{v^2-w^2}$ is shown in (f) (the contour interval in (d),(e),(f) is $2\times10^{-10}~h U_c$).
These  Reynolds stresses and roll circulations  emerge when a spanwise homogeneous field of fluctuations white in energy
with $k_x/\alpha=3$  is strained for only $0.001~h/U_c$ units of time 
by the low-speed streak of Fig. 
\ref{fig:ali2} centered at $z=0$.   
This figure shows that the universal structure of the Reynolds stresses supporting a streak emerges immediately through the straining of a
random homogeneous field of perturbations by the streak.}
 \label{fig:strain}
\end{figure*} 
 
 %{fig:Veq_dns_rnl_kx} 
 
 \section{Universality in structure  of the Reynolds stresses  arising from  fluctuations about the mean  streak }
 
We have seen  that the SSP is  primarily supported by  the Reynolds stresses of the first
10 streamwise wavenumbers (cf.  Fig. \ref{fig:Veq_dns_rnl_kx}). 
The time-mean Reynolds stresses  at these wavenumbers exhibit  a notable universality in structure 
about the  time-mean streak for the case of both the low-speed streak (Fig. \ref{fig:stresses_dns_low}, \ref{fig:stresses_10_12}) and  the high-speed  streak 
(Fig. \ref{fig:stresses_dns_high35}).
 Universality and self-similarity  of the time-mean structure of  fluctuations about time-mean flows has  been found to characterize wall-bounded turbulence
\citep{DelAlamo-etal-2006,Hwang-Cossu-2010b,Lozano-Duran-Jimenez-2014,Hwang-2015,Hellstrom-etal-2016}
and it has been demonstrated that this  property  derives from the linear interaction of the fluctuations
with the time-mean flow \citep{Farrell-Ioannou-1993b,DelAlamo-Jimenez-2006,Moarref-etal-2013,McKeon-2019,Vadarevu-etal-2019,Hwang-Eck-2020,Holford-Hwang-2023}.
Here we verify the universality of the time-mean  structure  of the large scale fluctuations that are  collocated with the low-speed and high-speed streak, which was already apparent
in Fig. \ref{fig:SVkx}, and attribute  the self-similarity  of these structures to the linear interaction of the fluctuations with  the time-mean streak.
The  typical structure of Reynolds stresses  of the fluctuations in NSE100 in low-speed streaks is shown in Fig.    \ref{fig:stresses_dns_low}, \ref{fig:stresses_10_12}  and 
in Fig  \ref{fig:stresses_dns_high35}
for  high-speed streaks. These figures show the universality in streamwise wavenumber of the structure of the
time-mean Reynolds stresses implying universality  of the mechanism sustaining and regulating the low speed streak. Remarkably, the universal structure
of the fluctuations on the streak will be shown to arise from the growth of fluctuations excited white in energy
 so that their structure arises solely from the optimal growth properties
of the streak and  does not require the introduction of color to the excitation.

The structure of the $\langle \overline{uw} \rangle$ (the second column of figures \ref{fig:stresses_dns_low}, \ref{fig:stresses_10_12} and  \ref{fig:stresses_dns_high35}) and of the  $\langle \overline{v^2-w^2} \rangle$ 
(the fourth column of figures \ref{fig:stresses_dns_low}, Fig. \ref{fig:stresses_10_12} and  \ref{fig:stresses_dns_high35})
Reynolds stress components are directly interpretable.
The structure of $\langle \overline{uw} \rangle$   
indicates that energy is being transferred in the mean  from the spanwise varying mean streak to the fluctuations.
This  
reflects the mechanism by which the fluctuations are sustaining while at the same time regulating the streaks. 
Near the centerline of a low speed streak $\partial_z \langle \overline{uw} \rangle <0$ indicating that on 
average the fluctuations are being sustained by gaining kinetic energy 
from the streak (cf. second column of Fig. \ref{fig:stresses_dns_low}). 
The opposite polarity of the $\langle \overline{uw} \rangle$  is found as required for sustaining the fluctuations in the case of a high speed streak (cf. second column of Fig. \ref{fig:stresses_dns_high35}).  
The $\langle \overline{v^2-w^2} \rangle $ Reynolds stress, identified as the asymmetric component of the Reynolds normal stress,  was shown 
above to be the primary source of roll acceleration supporting the streak through the lift-up mechanism.  
As discussed in the previous section the minimum of the normal stress at the centerline of
the low-speed streak indicates dominance of the $\cal S$ component of the fluctuations, 
consistent with the primacy of this term in providing
the  roll forcing
maintaining the low-speed
streak through lift-up (cf. Fig. \ref{fig:stresses_dns_low}, \ref{fig:stresses_10_12} (last column)).
In high-speed streaks the minimum of $\overline{v^2-w^2}$ occurs at the wings of the streak   (cf. Fig. \ref{fig:stresses_dns_high35}) 
indicating the dominance of the $\cal V$ fluctuations over the $\cal S$
 at the wings of the streak,  as is clear in Fig. \ref{fig:SVstresses_high} 
from the contribution to this stress from the ${\cal S}$ and $\cal V$ fluctuations separately. 

\begin{figure*}
\centering
\includegraphics[width=32pc]{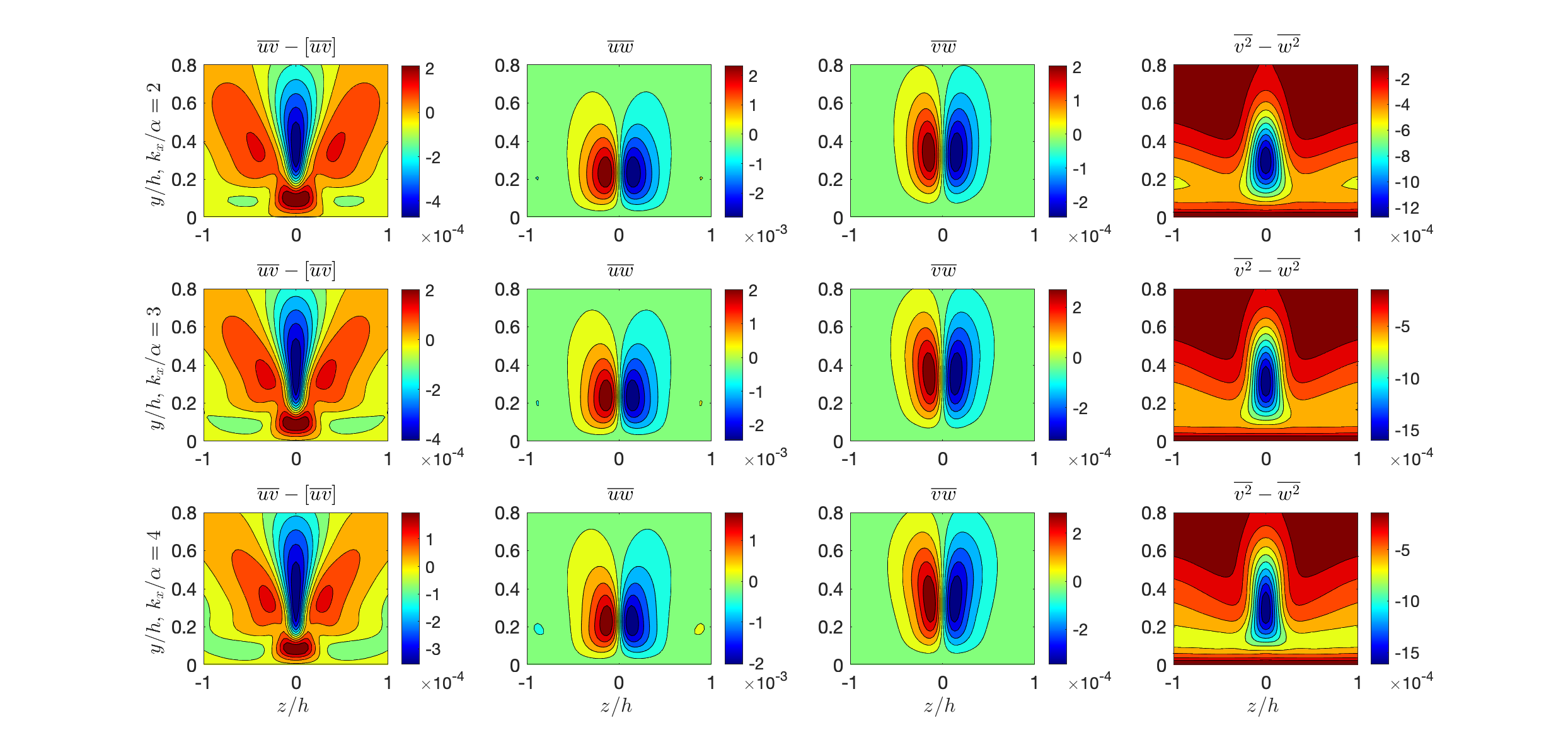}
\vspace{-1em}
\caption{As in Figure \ref{fig:stresses_dns_low} 
but showing the time-mean  Reynolds stresses  of fluctuations obtained using the 
$T_d=30 h/U_c$ STM covariance  resulting from stochastically exciting the time-mean
low-speed streak of NSE100 white in energy.  Results are shown 
for fluctuations with  $k_x/\alpha=2$ (top row), $k_x/\alpha=3$ (middle row)  
and $k_x/\alpha=4$ (bottom row).} \label{fig:stresses_STM_L}
\end{figure*}

 \begin{figure*}
\centering
\includegraphics[width=32pc]{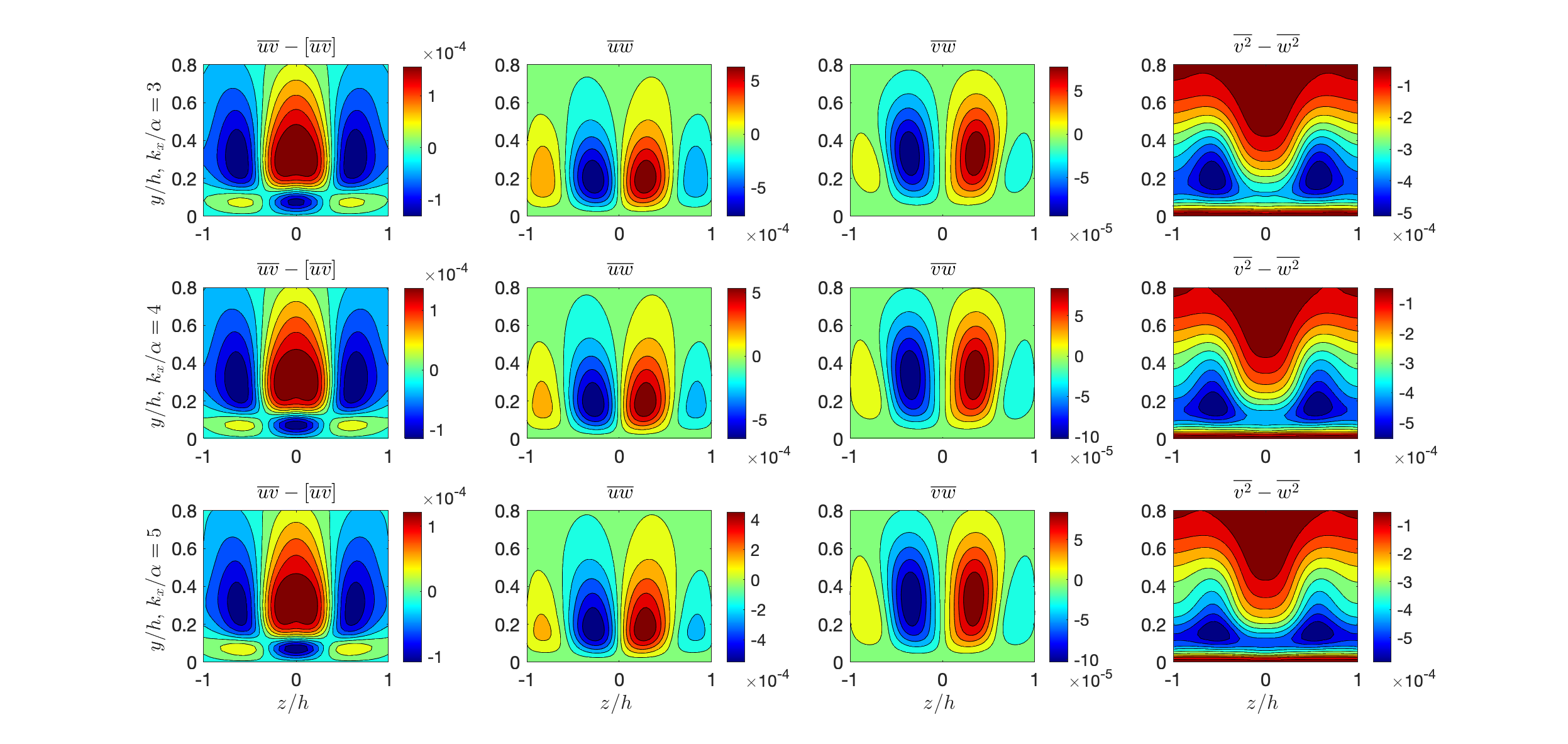}
\vspace{-1em}
\caption{As in Fig. \ref{fig:stresses_dns_high35}
%\ref{fig:stresses_STM_L} 
but showing the time-mean  Reynolds stresses  of fluctuations  obtained using 
the 
$T_d=30 h/U_c$ STM covariance  resulting from stochastically exciting the
high-speed streak of NSE100 white in energy. Results are shown for fluctuations with  $k_x/\alpha=3$ (top row), $k_x/\alpha=4$ (middle row)  and $k_x/\alpha=5$ (bottom row).} 
\label{fig:stresses_STM_H}
\end{figure*}

\section{Tracing the origin of the universality of the Reynolds stresses supporting the R-S to the optimal growth of perturbations on the R-S}

We have seen that there is a universal mechanism producing Reynolds stresses properly collocated to result in streak growth in turbulent shear flow.  
This   remarkable  result has precedence in earlier work in which it was
shown that even an infinitesimal streak perturbation
imposed on a random spanwise homogeneous field of turbulence in a shear flow organizes roll-inducing Reynolds 
stresses resulting in an unstable mode with R-S form arising from the streak perturbation \citep{Farrell-Ioannou-2012,Farrell-Ioannou-2017-bifur,Farrell-Ioannou-2022}.  
This result was ascribed to the structure of optimal perturbations which dominate
the perturbation variance as  a necessary 
consequence of the completeness of basis functions which requires that a sufficiently  random field will have a projection on every basis 
function and in a non-normal dynamics, such as a shear flow, only a small set of these projections grow appreciably over time in the energy norm.   
These structures can be identified using singular value decomposition to be the optimal perturbations.  
It follows that in the stochastic background field of shear turbulence a small set 
of optimal perturbations form a basis in the energy norm for the set of energy active fluctuations that
determine the  fluctuations obtaining significant amplitude.
The hypothesis to be tested is whether the optimal perturbations on a 
streak in a shear flow evolve correlated with the streak just so as 
to force the streak to grow by inducing 
roll forcing  that results in a streak-amplifying   lift-up process.  
The implication of this hypothesis being  verified is that, in the random field of the 
turbulent background, the set of growing structures that spontaneously develop  
are responsible for the universality of the Reynolds stresses and also that these optimal perturbations
tend to destabilize any perturbation with streak form  giving rise to a universal 
streak destabilizing mechanism  that is a general property of turbulence in shear flow.
%It has been demonstrated 

We test this unlikely hypothesis
by  calculating the ensemble mean covariance  of randomly excited perturbations imposed on a mean streak.
In this stochastic turbulence model (STM) the ensemble mean covariance,  $\C_{k_x}$, of the fluctuations with 
streamwise wavenumber $k_x$ (cf. Equation \eqref{eq:Ck}) 
that develop through the non-normal interaction with the mean flow, $U$, satisfy
the time-dependent Lyapunov equation
\begin{gather}
\frac{d \C_{k_x}}{ dt} = \A_{k_x}(U) \C_{k_x} + \C_{k_x} \A_{k_x}^{\dagger}(U) + \I ~,
\label{eq:dcdt}
\end{gather} 
which, it is useful to note, is  also the second  cumulant equation of S3T.
$\A_{k_x}(U)$ is the linear operator governing the evolution of the fluctuations
with wavenumber $k_x$ about the mean flow, $U$, $\I$ is the spatial covariance of the stochastic forcing, which is taken 
as the identity in order that all degrees of freedom are excited equally in energy, and $\dagger$ denotes the Hermitian transpose \citep{Farrell-Ioannou-1993e}.
 The mean flow $U \hat{x}$  considered is  hydrodynamically stable (it is our stable low or high speed streak).  While 
all fluctuations eventually decay, continual excitation
produces a finite covariance, which is  dominated by the structures that grow the most  by non-normal interaction 
with the mean flow over the interval of the development 
of $\C_{k_x}$. The dominant structures of the covariance and the associated Reynolds stresses are 
the   optimal perturbations with optimization taken over the time  chosen for the development of  $\C_{k_x}$. In  \cite{Nikolaidis-POD-2023} the dominant  POD modes 
of the covariance that develops in \eqref{eq:dcdt} in the background of the time-mean low-speed streak shown in Fig. \ref{fig:ali2}
were obtained. It was shown there that the dominant POD modes reflect the average structure of the optimal perturbations that grow on the streak
and consequently provide a characterization of the Reynolds stresses and of the induced roll forcing.

\begin{figure}
\centering
\includegraphics[width=0.75\columnwidth]{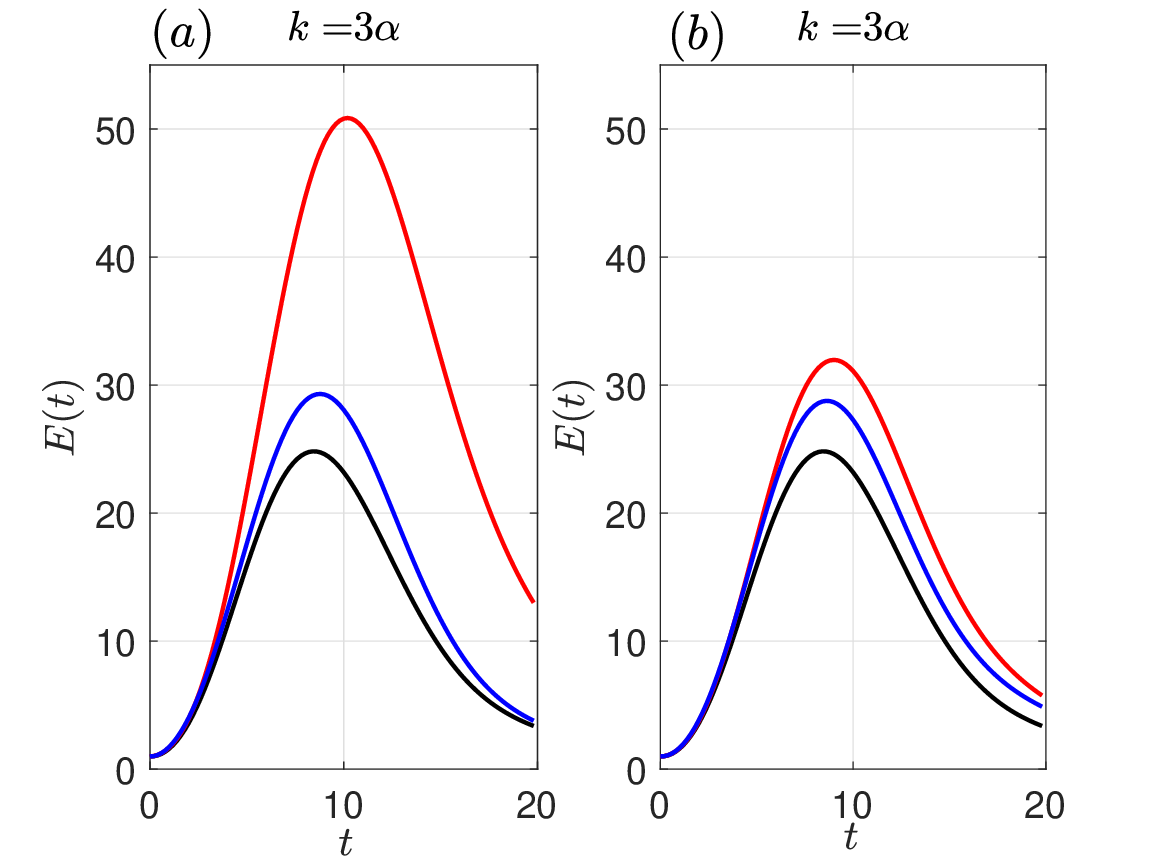}
\caption{Time evolution  of the energy of the $\cal S$ (red) and $\cal V$ (blue) $T=10~h/U_c$ optimal perturbations
in a flow with a low speed streak (a) and a high-speed streak with the same structure (b)  for streak amplitude $\varepsilon=1$.
The corresponding energy growth of the  ${\cal S}$ and $\cal V$   optimals   with 
no streak $\varepsilon=0$ are indicated with the black line, in this case the growth of 
the ${\cal S}$ and $\cal V$ optimal perturbations is equal.
This figure shows that the spanwise shear increases the  energy growth
of both ${\cal S}$ and $\cal V$ perturbations but that the low-speed streak supports substantially greater growth of the S optimal perturbation.
Perturbations have  $k_x/\alpha=3$.}
\label{fig:Et}
\end{figure} 

First  consider the covariances that develop  when the mean flow $U$ in \eqref{eq:dcdt} is  spanwise independent.
This to a good approximation occurs in the upper region, $y/h >1$, of the time mean flow in Fig. \ref{fig:ali2}.
In that case the ${\cal S}$ and $\cal V$ fluctuation components are equal on average modulo a spanwise shift, the $\C_{k_x}$ are  therefore
spanwise homogeneous  and the associated Reynolds stresses do not produce  roll forcing
as  $\langle \overline{v^2- w^2}\rangle$ is spanwise constant and 
 $\langle \overline{vw} \rangle=0$ (cf. equation \eqref{eq:Omega} and the streak acceleration in the $y/h>1$ region of Figures \ref{fig:DNS_low}a, \ref{fig:RNL_low}a and \ref{fig:DNS_high}a).
Upon introducing in \eqref{eq:dcdt} the mean flow of  the streak of Fig. \ref{fig:ali2}, indicated as $U_s$, the covariances
are no longer spanwise homogeneous
and the Reynolds stresses produce roll forcing.
Consider  \eqref{eq:dcdt}  integrated forward with the streak mean flow in Fig. \ref{fig:ali2} and initial condition the spanwise homogeneous equilibrium covariance, $\C_{hom,k_x}$,
that emerges asymptotically when the mean flow is the spanwise independent time-mean flow.  The
inhomogeneous covariance $ \C_{inh, k_x}$ that will develop according to \eqref{eq:dcdt} in time $\delta t$ after the introduction of the  streak  is:
\begin{gather}
\C_{inh,k_x} =(\A_{k_x}(U_s) \C_{hom,k_x} + \C_{hom,k_x} \A_{k_x}^{\dagger}(U_s) ) \delta t ~.
\label{eq:dcdt1}
\end{gather} 
 The Reynolds shear and normal stresses produced by $\C_{inh,k_x}$
are shown in Fig. \ref{fig:strain}a,b after  integration of  \eqref{eq:dcdt} for  $\delta t=0.001~h/U_c$ units of time. The
roll-circulation  induced through the straining of the field over this short interval of time  is shown in Fig. \ref{fig:strain}c,d,e,f.  
Remarkably, the  universal structure of the Reynolds stresses and of the  roll forcing  seen in the time mean statistics of NSE100  and RNL100 manifests 
instantaneously upon the introduction of the streak in the flow.
This indicates that straining over an infinitesimal time interval of a spanwise  homogeneous fluctuation field   by a low-speed streak
favors the $\cal S$ component of the field over the $\cal V$ producing correctly configured roll forcing to destabilze the streak.  

Note that because $\A_{k_x}(U)$ depends linearly on $U$,  reversing the sign of $U_s$ in Equation \eqref{eq:dcdt1} 
reverses the sign of all the Reynolds stresses  and Equation \eqref{eq:dcdt1} predicts that
a high-speed streak strains a spanwise homogeneous field of turbulence to produce push-down 
 reinforcing the high-speed streak.  
Because changes of the strength of the streak in \eqref{eq:dcdt1}  are associated only with changes in the time scale,
 these results also apply to infinitesimal streaks and it is in fact  the mechanism underlying
the  {exponential  growth of the R-S}   which, while clearly manifested in DNS and RNL, has analytic expression { as a modal instability} only in the 
infinite ensemble framework of S3T theory 
\citep{Farrell-Ioannou-2017-bifur}. 
This example application underscores the fundamental theoretical importance of 
analyzing  the dynamics of an infinite ensemble
of realizations and the value of convincingly demonstrating the dynamical similarity
among DNS, RNL and S3T in order to exploit  the power afforded by the analytic structure of S3T to understand NSE turbulence.

Having shown that infinitesimal straining of a spanwise homogeneous field of 
turbulence by a low or high-speed streak produces R-S 
destabilizing Reynolds stresses, 
we turn next to straining a turbulent field over a time interval typical of 
fully developed turbulence by considering the Reynolds stresses that develop in the STM over a period of  $T_d=30 h/U_c$
initiated with the asymptotic covariance in the absence of a streak.
This period is selected because it is the typical coherence time of integral scale fluctuations in the turbulent flow (cf. \cite{Lozano-Duran-etal-2021}).
The Reynolds stresses that are obtained by this STM, shown in Fig. \ref{fig:stresses_STM_L} (cf. Fig. \ref{fig:stresses_dns_low}) for the low-speed streak  and  Fig. \ref{fig:stresses_STM_H}   (cf. Fig. \ref{fig:stresses_dns_high35}) for the high-speed streak, verify that 
non-normal linear interaction 
between the streak and white in energy  random perturbations give rise to average perturbations producing the roll-forcing 
 ensemble mean Reynolds stresses observed in DNS.  
 
 Note that in the case of
 infinitesimal straining the structure of the stress distribution  in high-speed streaks  is identical to  that in low speed streaks except for
 a change in sign. 
 In contrast, the stress distributions in finite amplitude high and low speed streaks differ substantially
 in structure, as seen in Fig. \ref{fig:stresses_STM_L}   and  Fig. \ref{fig:stresses_STM_H}.  Nevertheless, in both cases the stress distributions 
 induce roll forcing  that maintain 
 the imposed streak.  The difference in the stress distribution between low and high-speed  finite amplitude streaks results from
 differences in the optimal perturbation growth  in low and high-speed streaks,
 which  favors the growth of   
  S optimal perturbations in low-speed  streaks, as was previously  noted  by \cite{Hoepffner-etal-2005}.
   
 We illustrate in Fig. \ref{fig:Et} this divergent behavior of the ${\cal S}$ and $\cal V$ optimals in
 the presence of the time-mean flows $U_m(y)\pm \varepsilon U_s(y,z)$, where $U_m(y)$ is the spanwise mean flow
 and  $U_s(y,z)$ is the time mean low-speed streak of Fig. \ref{fig:ali2} with streak amplitude
$\varepsilon=0,\pm0.4,\pm1$.  
This figure shows  that  the optimal perturbation growth  increases as the amplitude of the streak increases and that  the increase is substantial  
when the streak is low-speed and marginal 
when the streak is high-speed. The  optimization time $T=10~h/U_c$  was chosen to correspond to the global optimal time.
Energy transfer from the mean spanwise shear to the perturbations, $-\int_{\cal D} dy dz~ \overline{uw} U_z$,
is the energy source that accounts for the increased perturbation growth in the presence of the streak, and especially so when a low-speed streak is present because
flows with low-speed streaks have a relatively smaller wall-normal shear and the perturbations are less readily sheared over by the wall-normal shear, which limits their potential growth.
Differences  in the growth of perturbations in flows of
the form  $U_m(y)\pm \varepsilon U_s(y,z)$ is expected, because the flows are 
not  mirror images of each other.   

The pronounced asymmetry in the  growth of optimal perturbations in low-speed 
and high-speed streaks is surprising and has dynamical implications.
It implies, as we have seen reflected in the time-mean statistics of the DNS and also RNL,
that high-speed streaks are supported weakly by their Reynolds stresses, which contributes to  
the dominance of low-speed streaks in wall-bounded turbulence.

 \section{Conclusions}

In this work we have examined the dynamics  supporting the R-S 
in  plane Poiseuille turbulence at  $R=1650$ and verified that this dynamics  is substantially the same in RNL and DNS
and that it is the mechanism of R-S destabilization 
by transiently growing structures identified in S3T dynamics \citep{Farrell-Ioannou-2012}.  
Transient growth is shown to destabilize an imposed steak immediately as the background turbulence is strained by 
the streak and to continue to amplify the streak as optimally growing 
structures contained in the background turbulence evolve over finite time,
as is required for both initial destabilization of the R-S and its maintenance at finite amplitude.

In order to study the R-S  formation, maintenance and regulation to its finite amplitude equilibrium we have departed from 
the traditional decomposition of wall-bounded turbulence into  time-mean and   fluctuation fields, in which the 
R-S is relegated to comprise a part of the
fluctuation field.
We have rather chosen a  streamwise mean  decomposition because this 
partition  results in a SSD that comprises the fundamental dynamics 
of wall-turbulence in a  transparent manner. With the R-S contained in the mean flow
we  obtain a  second-order closure, referred to as S3T \citep{Farrell-Ioannou-2012}, %{Lozano-Duran-etal-2021},
that concisely captures the structure and dynamics  of wall-turbulence.

The validity and utility of S3T theory is evident 
from the fact that it provides the means for the  analytic study 
of the stability  of the attractors of the SSD of turbulent shear flows. 
Consider as example a plane wall-bounded  flow with a statistical mean equilibrium profile  consistent with
an externally supplied spanwise homogeneous
field of  random fluctuations (a field of free-stream turbulence).  Application of S3T perturbation stability analysis reveals
  that this spanwise-uniform  mean flow and its associated 
fluctuation cumulant is { modally} unstable at large enough Reynolds numbers,  giving rise to a mean flow that includes
rolls and streaks \citep{Farrell-Ioannou-2017-bifur}.  
That this fundamental symmetry breaking instability has analytic expression only in S3T,
while being clearly manifest in both DNS and RNL,  
indicates the analytic utility of adopting  S3T  for the study of turbulence in shear flow.
This point of view is implicitly adopted in the classical picture of the SSP cycle \citep{Hamilton-etal-1995}, which involves a 
self-sustaining quasi-linear interaction of the streamwise mean  with the fluctuations,
as analytically embodied in the S3T/RNL dynamics. 
Because S3T and RNL  have the same dynamical structure (RNL is essentially S3T with the second cumulant approximated by a finite
ensemble)  we can take 
RNL simulations as confirming at higher Reynolds numbers than S3T can be integrated
that this  quasi-linear interaction, 
with this  definition of the mean, produces, within the framework of the Navier-Stokes equations,
a sustained  SSP cycle and realistic turbulent states 
\citep{Thomas-etal-2014,Bretheim-etal-2015,Farrell-etal-2016-VLSM,Farrell-etal-2016-PTRSA}.
The key ingredient of the SSP cycle, as  identified in \citep{Farrell-Ioannou-2012} and extensively verified in RNL simulations, 
is that  in the presence of a streak the non-normal growth of fluctuations results in
Reynolds stresses that drive roll circulations that  reinforce  the pre-existing
streaks in the flow. This  is also the underlying mechanism of the S3T { modal}  instability 
discussed in \citep{Farrell-Ioannou-2012,Farrell-Ioannou-2017-bifur}: any 
flow perturbation with streak form induces  
ensemble fluctuation Reynolds stresses that   lead to   collocated roll circulations that, at high 
enough Reynolds numbers, lead 
to  exponential growth of the R-S.  
In this work we began by verifying  that the turbulent fluctuations in DNS and in RNL become configured in the presence of a streak so as to 
induce  roll circulations that reinforce  pre-existing
streaks in the flow. While roll forcing by fluctuation Reynolds stresses was previously identified and verified to be the mechanism of R-S formation
in  S3T, RNL and DNS, the exact dynamical mechanism producing the required collocated roll forcing was left unidentified.
In this work we showed using data from a DNS  that in the time mean the ${\cal S}$ and $\cal V$ components of the fluctuations are
linearly statistically independent and it is the $\cal S$  fluctuations about the centerline of the streak
that produce roll circulations  leading  to lift-up  in shear regions, 
strengthening pre-existing  low-speed streaks and weakening high-speed streaks,
while it is  the $\cal V$ fluctuations that produce the opposite effects resulting in amplification of high-speed streaks.  
In a homogeneous turbulent background field and without a perturbation of streak form these opposing streak 
forming tendencies  cancel exactly leading
to no roll formation. However, in the presence of a   streak,
 exact stress  balance between the $\cal S$  and $\cal V$ components is 
 disrupted  so that  for a low-speed streak the $\cal S$
roll-forming  stresses dominate over the $\cal V$ roll-destroying stresses while the  opposite is true in  high-speed streaks. 
In this way the presence of a  streak partitions the fluctuation stresses between  ${\cal S}$ and $\cal V$ components 
in just the manner required for its amplification.
While both the ${\cal S}$ and $\cal V$ fluctuations are present in both low-speed and high-speed streaks, 
so that e.g. careful data analysis would reveal $\cal V$ structures consistent 
with hairpin vortices coincident with low-speed streaks, we show  in this work that these 
V structures oppose rather than support the low-speed streak.

When diagnosis of the streamwise varying fluctuation Reynolds stresses is made we find
 that the Reynolds stress that dominates in the formation and maintenance of the R-S 
is the asymmetry in the Reynolds normal stress, $\langle \overline{v^2-w^2} \rangle $,  and primarily of the $\langle \overline{w^2} \rangle$
 component that develops due to the asymmetric non-normal amplifications of the 
 ${\cal S}$ and $\cal V$ components of the fluctuations arising  in the turbulent field,  and that the distribution of this 
normal stress determines the direction of the roll circulation. In a forthcoming publication we explain how 
this normal stress distribution determines the direction of the roll forcing \citep{Farrell-Ioannou-2022}.
This remarkable identification of  the primary role of the 
asymmetric Reynolds normal stress in the dynamics of the R-S  points to a novel interpretation
of the origin of this structure that underlies the maintenance of wall-turbulence.
The utility of  verifying that  the  same mechanism supports wall-turbulence in the 
three representations of NS dynamics, the S3T SSD closure, the RNL approximation of the S3T SSD closure  and DNS, 
lies in the fact that the S3T is analytically complete in  the dynamics of its turbulence 
whereas the DNS has proven recalcitrant to reveal its fundamental dynamics.  The S3T/RNL system
being both analytically transparent
and numerically tractable provides a powerful tool for understanding the fundamental dynamics of wall-turbulence.  In addition to its theoretical utility, 
the quasi-linear structure of S3T/RNL promises to allow extension of
the powerful methods of linear control 
to 
address other problems  associated with both understanding and controlling turbulence in shear flow.

 \section*{Declaration of interest}

The authors report no conflict of interest

\bibliographystyle{jfm}
%\bibliography{../../bibfile/basic_references}

\end{document}